\documentclass{IEEEtaes}
\usepackage{color,array,amsthm}
\usepackage{amsmath,amsfonts,amssymb}
\usepackage{graphicx}
\usepackage{algorithmic}
\usepackage{epsfig,algorithm}
\usepackage{acronym}
\usepackage{textcomp}
\usepackage{stfloats}
\usepackage{tabularx}
\usepackage{xcolor}
\usepackage{url}
\usepackage{verbatim}
\usepackage{multirow}
\usepackage{float}
\usepackage{soul}
\usepackage{cancel}
\usepackage{cite}
\usepackage{balance}
\usepackage{hhline}
\usepackage[T1]{fontenc}
\usepackage[utf8]{inputenc}

\usepackage[caption=false,font=normalsize,labelfont=sf,textfont=sf]{subfig}
\def\BibTeX{{\rm B\kern-.05em{\sc i\kern-.025em b}\kern-.08em
T\kern-.1667em\lower.7ex\hbox{E}\kern-.125emX}}

\newtheorem{theorem}{Theorem}
\newtheorem{proposition}{Proposition}
\newtheorem{corollary}{Corollary}

\newtheorem{remark}{Remark}

\jvol{XX}
\jnum{XX}
\jmonth{XXXXX}
\paper{1234567}
\pubyear{2025}
\doiinfo{TAES.2025.Doi Number}
\input{AcronymsListFinal}
\acresetall

\begin{document}

\title{An HFM-Inspired Random Access Preamble Design for NTN under High Doppler}

\author{MEHWISH BIBI}
\affil{Istanbul Medipol University, Istanbul, T\"{u}rkiye}

\author{SAIRA RAFIQUE}
\affil{Vestel Electronics, Manisa, T\"{u}rkiye}

\author{AHMED NAEEM}
\affil{National University of Sciences and Technology, Islamabad, Pakistan}

\author{H\"{U}SEYIN ARSLAN}
\member{Fellow, IEEE}
\affil{Istanbul Medipol University, Istanbul, T\"{u}rkiye}

% \receiveddate{Manuscript received XXXXX 00, 0000; revised XXXXX 00, 0000; accepted XXXXX 00, 0000.\\
% This work was supported in part by [funding source and grant number].%}

% \corresp{{\itshape (Corresponding author: Mehwish Bibi.)}}

\authoraddress{Mehwish Bibi and H\"useyin Arslan are with the Department of Electrical and Electronics Engineering, Istanbul Medipol University, Istanbul 34810, T\"urkiye (e-mail: \href{mailto:mehwish.bibi@std.medipol.edu.tr}{mehwish.bibi@std.medipol.edu.tr}; \href{mailto:huseyinarslan@medipol.edu.tr}{huseyinarslan@medipol.edu.tr}). Saira Rafique is with the IPR and License Agreements Department, Vestel Electronics, 45030 Manisa, T\"urkiye (e-mail: \href{mailto:saira.rafique@vestel.com.tr}{saira.rafique@vestel.com.tr}). Ahmed Naeem is with the School of Electrical Engineering and Computer Science, National University of Sciences and Technology, Islamabad, Pakistan (e-mail: \href{mailto:ahmed.naeem@seecs.edu.pk}{ahmed.naeem@seecs.edu.pk}). \\\textcolor{red}{\textbf{(This work has been submitted to the IEEE for possible publication. Copyright may be transferred without notice, after which this version may no longer be accessible.)}}}

\markboth{BIBI ET AL.}{HFM-BASED RANDOM ACCESS FOR ROBUST SYNCHRONIZATION IN LEO-NTN}

\maketitle
%%%%%%%%%%%%%%%%%%%%%%%%%%%
\begin{abstract}
Non-terrestrial networks (NTNs) are a key enabler of ubiquitous 6G connectivity, but the high orbital velocity and long propagation distances in low-Earth orbit (LEO)-NTN operation introduce large Doppler shifts and substantial delay uncertainty that challenge New Radio (NR) physical random access channel (PRACH) design. Conventional Zadoff–Chu (ZC) and linear frequency modulated (LFM) preambles are particularly vulnerable, as Doppler-induced ambiguity and delay–Doppler coupling degrade timing estimation and preamble identification. This paper proposes a hyperbolic frequency modulation (HFM)-inspired PRACH preamble for robust synchronization and reliable identification under uncompensated or unknown Doppler, detected with a conventional matched-filter receiver so that the gains reflect the preamble design. A unified delay–Doppler ambiguity function framework characterizes the self- and cross-ambiguity behavior of ZC, LFM, and HFM-inspired preambles, and a scaling-factor-based codebook ensures multi-user separability. Simulation results under NTN channel conditions confirm higher detection probability, lower timing root-mean-square error (RMSE), and improved peak-to-sidelobe and integrated sidelobe levels compared with ZC and frequency-domain superposed LFM baselines.
\end{abstract}

%%%%%%%%%%%%%%%%%%%%%%%%%%%
\IEEEpeerreviewmaketitle
% \vspace{-3mm}
\section{Introduction}
\label{sec:introduction}

In beyond-5G and 6G systems, \acp{NTN} complement terrestrial infrastructure with \ac{LEO} satellites, high-altitude platforms, and aerial nodes to deliver ubiquitous connectivity. \ac{LEO} constellations at altitudes of approximately $300$–$600$~km offer a favorable balance between coverage, latency, and deployment cost~\cite{11010845}, extending broadband, remote-area, maritime, aeronautical, and critical-infrastructure links to regions where terrestrial coverage is sparse or economically infeasible~\cite{10492466,9650576,11363335}. Accordingly, \acp{NTN} are introduced in the \ac{3GPP} \ac{NR} framework in Release 17, and requirements of physical-layer procedures, including synchronization, random access, and timing advance under large delay and Doppler uncertainty, are discussed in subsequent releases~\cite{3gpp-tr-38.811, 3gpp-tr-38.821}. The \ac{PRACH} is central to these procedures, providing initial access~\cite{10947167}, uplink timing alignment, and preamble identification under significant channel impairments.

In \ac{LEO}-based \ac{NTN}, the \ac{PRACH} procedure is challenged by large Doppler offsets, high Doppler rates from satellite motion, and long propagation delays~\cite{7962769, 9978044}, which violate the quasi-static channel assumptions underlying terrestrial \ac{NR} \ac{PRACH} design. The standard \ac{ZC} preambles used in \ac{LTE} \ac{PRACH} rely on cyclic autocorrelation properties preserved only in the absence of frequency offset; under residual Doppler, \ac{ZC} sequences suffer correlation-peak displacement that degrades timing accuracy and produces ambiguous preamble identification~\cite{11195401,alarcon2026cfo}. This degradation worsens in multi-user random access, where simultaneous transmissions increase cross-correlation leakage~\cite{8286861, 10720128}, and in wide \ac{LEO} beams, whose large footprint aggregates many access attempts and intensifies preamble collisions and congestion~\cite{chen2026channelaware}.

Reliable \ac{ZC}-based \ac{PRACH} in \ac{NTN} relies on \ac{UE}-side Doppler and timing pre-compensation from \ac{GNSS} and satellite ephemeris data~\cite{3gpp-tr-38.811, 3gpp-tr-38.821}. However, \ac{GNSS} is unavailable in indoor or obstructed environments and impractical for terminals lacking reliable positioning or processing resources. Even with pre-compensation, residual Doppler can remain large enough to invalidate the \ac{ZC} correlation properties, motivating \ac{GNSS}-resilient waveforms and receivers that do not depend on accurate \ac{UE}-side pre-compensation~\cite{georganaki2026mirage}.

Several works address \ac{ZC} Doppler sensitivity through modified \ac{ZC}-based \ac{PRACH} preambles~\cite{9681944,10001203,10511361,11173866}, using root selection, block-cyclic constructions, extended cyclic prefixes, and zero-correlation-zone designs to enlarge the preamble pool and improve detection under Doppler and delay spread~\cite{jeong2026isc}. These approaches nonetheless inherit the lattice-like \ac{ZC} \ac{AF}, whose multiple comparable cross-ambiguity peaks in the delay--Doppler plane limit \ac{PID} separability under residual Doppler and multi-user interference.

Owing to their Doppler robustness in radar and sensing, \ac{LFM} chirp waveforms have been investigated as \ac{PRACH} candidates for high-mobility environments~\cite{9340403}, and the related \ac{AFDM} and \ac{DAFT}-domain preambles exploit chirp subcarriers to absorb delay and Doppler into a compact affine representation while enlarging the preamble pool~\cite{xu2026afdm,li2026dualoffset}. Although \ac{LFM} preserves strong matched-filter responses under frequency offset, its deterministic delay--Doppler coupling yields ridge-shaped \acp{AF}; slope mismatch alone therefore does not guarantee \ac{PID} separability, since cross-ambiguity peaks can approach the self-ambiguity peak along the affine ridge~\cite{balleri2017ambiguity}.

To improve \ac{LFM} robustness at higher Doppler, \ac{FDS-LFM} preambles superpose multiple short \ac{LFM} subsequences via frequency-domain phase rotations combined through an \ac{IDFT}, embedding \ac{PID} through structured phase offsets and detecting via subsequence-level correlation combining~\cite{10892192}. This improves missed-detection performance over \ac{ZC}-based \ac{PRACH} under large residual Doppler, but the underlying \ac{LFM} structure still produces extended cross-ambiguity ridges that constrain multi-user \ac{PID} reliability.

\Ac{HFM} waveforms offer a fundamentally different chirp structure, widely studied in radar, sonar, and underwater acoustics for their Doppler accuracy~\cite{balleri2017ambiguity,10810361,9691900, 11357567}. Their nonlinear instantaneous frequency reshapes the \ac{AF} geometry, suppressing the ridge and lattice formations of \ac{LFM} and \ac{ZC} while sharpening mainlobe localization and timing accuracy under severe Doppler~\cite{balleri2017ambiguity,10810361}, making \ac{HFM} well-suited to the unavoidable large Doppler and delay uncertainty of the \ac{NTN}-\ac{PRACH} procedure.

Despite these favorable characteristics, \ac{HFM} has not been systematically applied to \ac{NR}-compatible \ac{PRACH} design; its implications for \ac{PRACH} cross-ambiguity behavior, preamble identification, multi-user separability, and detection-theoretic metrics such as false-alarm and detection probability remain largely unexplored. This paper addresses these gaps through an \ac{HFM}-inspired \ac{PRACH} design for \ac{LEO} \ac{NTN} under large uncompensated or unknown Doppler.

%%%%%%%%%%%%%%%%%%%%%%%%%%%%%
\subsection{Contributions}
This paper addresses the \ac{PRACH} \emph{preamble design} problem for \ac{LEO}-\ac{NTN} random access under large unknown or uncompensated Doppler. We propose an \ac{HFM}-motivated preamble that exploits the inherent Doppler tolerance of \ac{HFM} signals within an \ac{NR}-compatible frame structure, and we evaluate it using a conventional matched-filter \ac{PRACH} receiver so that the observed gains are attributable to the preamble design itself rather than to any receiver-side processing. The main contributions of this work are summarized as follows:
\begin{itemize}
    \item We propose a novel \ac{HFM}-inspired \ac{PRACH} preamble for \ac{LEO}-\ac{NTN} operation under uncompensated or unknown Doppler. The design replaces only the frequency-domain sequence of the \ac{NR} \ac{PRACH}, leaving the numerology, framing, and receiver chain unchanged, and directly addresses the timing-bias and \ac{PID} separability limitations of conventional \ac{ZC}- and \ac{LFM}-based preambles without UE-side Doppler pre-compensation.
    \item A scaling-factor-parameterized \ac{HFM} codebook is introduced to support multiple \ac{UE} identities, with a derived minimum-spacing condition that keeps cross-ambiguity sidelobes strictly below the self-ambiguity peak across the delay--Doppler plane.
    \item We analytically characterize the self- and cross-ambiguity properties of \ac{ZC}, \ac{LFM}, and \ac{HFM} preambles within a unified framework, establishing that the \ac{HFM}-inspired preamble suppresses the lattice and ridge structures inherent to \ac{ZC} and \ac{LFM} and yields timing estimation free of the deterministic Doppler-dependent bias exhibited by \ac{ZC} and \ac{LFM}.
    \item Using the conventional receiver, we validate the analysis under \ac{3GPP}-compliant \ac{NTN} conditions, showing that the proposed preamble achieves higher detection probability, lower timing \ac{RMSE}, and improved \ac{PSLR}/\ac{ISLR} than \ac{ZC} and \ac{FDS-LFM} baselines in both single- and multi-user scenarios, confirming that the improvements originate from the preamble design alone.
\end{itemize}
%%%%%%%%%%%%%%%%%%%%%%%%%%%%%
\subsection{Paper Structure}
The remainder of this paper is organized as follows. Section \ref{Sec:system_model} presents the \ac{LEO}-\ac{NTN} system model, alongside the propagation impairments and the proposed \ac{HFM}-inspired \ac{PRACH} preamble. Section \ref{Sec_III} describes the receiver-side processing for timing estimation and \ac{PID}. Section \ref{Sec:performance_analysis} develops the analytical framework, where the self- and cross-ambiguity behavior is examined together with the false-alarm and detection probabilities. Simulation results are presented in Section \ref{sec:results}, where the proposed scheme is compared against \ac{ZC} and \ac{FDS-LFM} baselines, and Section \ref{sec:conclusion} concludes the paper.
%%%%%%%%%%%%%%%%%%%%%%%%%%%%%

%%%%%%%%%%%%%%%%%%%%%%%%%%%%%%%%%%%%%%%%%%
\begin{figure*} 
\centering 
\resizebox{1.9\columnwidth}{!}{
\includegraphics{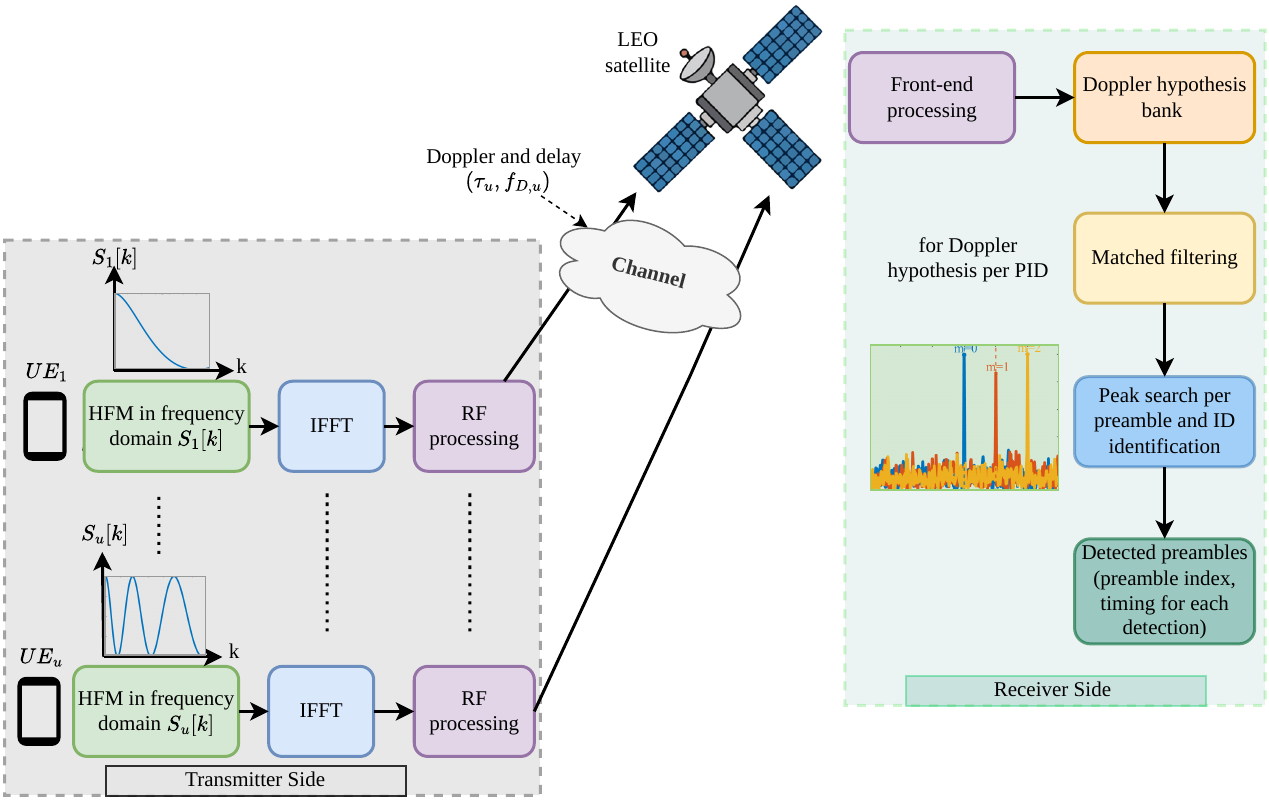}}
\caption{System model.}
\label{fig:Systemmodel}
\end{figure*}
%%%%%%%%%%%%%%%%%%%%%%%%%%%%%%%%%%%%%%%%%%

\section{System Model}

\label{Sec:system_model}
This section establishes the system model and proposed \ac{PRACH} preamble design for \ac{LEO} \ac{NTN} uplink \ac{RAP}. We consider a regenerative \ac{LEO} satellite deployment where \acp{UE} transmit \ac{PRACH} preamble without open-loop pre-compensation to initiate network access and establish uplink synchronization. The received signal is jointly impaired by large propagation delays, severe Doppler shifts, and multi-user interference, collectively challenging conventional \ac{ZC}-based \ac{NR} \ac{PRACH} designs. 

\subsection{NTN Uplink Scenario and Propagation Model}

We consider an uplink \ac{NTN} scenario in which \acp{UE} access the network through a satellite deployed at an orbital altitude of $a$. Each satellite is equipped with a regenerative payload that supports full \ac{NR} \ac{RAN} functionality, enabling uplink signals transmitted by the \acp{UE} to be received and processed directly onboard. The considered system focuses on the uplink \ac{RAP}, where \acp{UE} initiate network access by transmitting \ac{PRACH} preambles toward the serving satellite, as illustrated in Fig.~\ref{fig:Systemmodel}.

At \ac{LEO} orbital altitudes, the large satellite-to-\ac{UE} separation and high orbital velocity introduce propagation impairments substantially more severe than those encountered in terrestrial deployments. The one-way propagation delay between \ac{UE} and the serving satellite is given by
\begin{equation}
\tau = \frac{R}{c},
\end{equation}
where $R$ denotes the slant range and $c$ is the speed of light. At \ac{LEO} altitudes, $\tau$ reaches several milliseconds orders of magnitude larger than typical terrestrial values. Furthermore, the relative motion between the satellite and \ac{UE} further induces a Doppler frequency shift on the received signal, expressed as
\begin{equation}
f_{D} = \frac{v_r}{\lambda_c},
\end{equation}
where $v_r$ denotes the radial velocity component between \ac{UE} and the satellite, and $\lambda_c$ is the carrier wavelength. In \ac{LEO} \ac{NTN} scenarios, $f_{D}$ can reach several kilohertz, far exceeding Doppler shift values in terrestrial networks. The uncompensated $\tau$ and $f_{D}$ manifest as unknown impairments at the satellite receiver, where timing estimation and preamble identification are to be performed under large delay uncertainty and severe Doppler offset.

\subsection{Received Signal Model}
The \ac{NTN} service link between each $u$-th \ac{UE} and the satellite is modeled as a narrowband Rician fading channel, reflecting the dominant \ac{LoS} propagation geometry that arises from high satellite elevation angles and limited local scattering in \ac{LEO} \ac{NTN} scenarios~\cite{3gpp_tr_38811_r16, 10720128, 3gpp-tr-38.821}, with the Rician $K$-factor capturing the ratio of \ac{LoS} to scattered signal power and the coefficient $h_u$ aggregating the specular and diffuse contributions. The time-varying channel between $u$-th \ac{UE} and the \ac{LEO} satellite imposes a propagation delay $\tau_u$ and a Doppler shift $f_{D,u}$ on the transmitted waveform, yielding the received signal contribution at the satellite as
\begin{equation}
r_u(t) = h_u\, s_u\!\left(t - \tau_u\right) e^{j2\pi f_{D,u} t},
\end{equation}
where $s_u(t)$ denotes the transmitted \ac{PRACH} preamble waveform. The received signal at the satellite during a \ac{PRACH} occasion is the superposition of preamble transmissions from all $\mathcal{U}$ concurrently active \acp{UE}, and is expressed as
\begin{equation}
\label{received_eq}
r(t) = \sum_{u \in \mathcal{U}} h_u\, s_u\!\left(t - \tau_u\right) e^{j2\pi f_{D,u} t} + w(t),
\end{equation}
where $\mathcal{U}$ denotes the set of active \acp{UE} transmitting within the same \ac{PRACH} occasion and $w(t) \sim \mathcal{CN}(0,\sigma^2)$ represents additive white Gaussian noise with variance $\sigma^2$. The signal model in \eqref{received_eq} reveals that the satellite receiver must perform timing estimation and preamble identification under the joint presence of $\tau_u$, $f_{D,u}$, and structured multi-user interference. These impairments collectively undermine the correlation properties of conventional \ac{ZC}-based \ac{PRACH} preambles and necessitate a Doppler-resilient preamble waveform capable of reliable operation under the propagation conditions inherent to \ac{LEO} \ac{NTN} scenarios.

\subsection{Proposed HFM-Inspired PRACH Preamble}

In \ac{LEO} \ac{NTN} scenarios, the large uncompensated $f_{D}$ destroys the cyclic orthogonality of conventional \ac{ZC} sequences employed in \ac{NR} \ac{PRACH}. Specifically, a carrier-frequency offset applied to a \ac{ZC} sequence displaces the matched-filter correlation peak away from the true propagation delay, thereby coupling timing estimation with frequency uncertainty and degrading both preamble detection and identification. A preamble waveform whose correlation peak remains at the true delay without the deterministic Doppler-induced shift of \ac{ZC} is therefore required.

The \ac{HFM} waveform satisfies this requirement. Owing to its nonlinear logarithmic phase evolution, a carrier frequency offset applied to an \ac{HFM} signal does not displace the correlation peak from the true delay, attenuating it instead. Consequently, timing estimation is decoupled from Doppler uncertainty without any \ac{UE}-side frequency pre-compensation. This property is established formally in the performance analysis of Section~\ref{Sec:performance_analysis}.

The proposed preamble retains the standard \ac{NR} \ac{PRACH} transmission structure, comprising a \ac{CP} of length $N_{\text{CP}}$, a preamble sequence of length $N_{\text{SEQ}}$, and a \ac{GP} of length $N_{\text{GP}}$, with total duration
\begin{equation}
  N_{\text{PRACH}} = N_{\text{CP}} + N_{\text{SEQ}} + N_{\text{GP}}.
\end{equation}
Only the preamble sequence itself is replaced; the \ac{PRACH} numerology, subcarrier allocation, and framing remain unchanged.

\subsubsection{Proposed Preamble Sequence Definition}
We propose a frequency-domain \ac{PRACH} preamble sequence whose phase evolves logarithmically with the $k$ index, leveraging the Doppler-tolerance property of \ac{HFM}. The logarithmic phase of the proposed sequence decouples delay and Doppler estimation by preserving the correlation peak at the true $\tau$ under $f_D$. The sequence is defined in closed form directly on the \ac{OFDM} subcarrier grid, satisfies $|S[k]|=1\;\forall\,k$, and is compatible with the \ac{NR} \ac{PRACH} signal generation chain of 3GPP~TS~38.211~\cite{3gpp-ts-38.211}. The \ac{HFM} preamble sequence
of length $N_{\text{SEQ}}$ is defined as~\cite{10810361}
\begin{equation}\label{tx_eq1}
  S[k] =
  \exp\!\left(
    j\,\frac{2\pi}{\beta_{\text{HFM}}}
    \ln\!\left(1 + \frac{\beta_{\text{HFM}}\, k}{N_{\text{SEQ}}}\right)
  \right), \quad k = 0, \dots, N_{\text{SEQ}}-1,
\end{equation}
where $\beta_{\text{HFM}} = (f_L - f_H)/(f_L\, f_H\, T_{\text{SEQ}})$ is the \ac{HFM} rate parameter, $f_L$ and $f_H$ denote the lower and upper edges of the allocated \ac{PRACH} bandwidth, respectively, and $T_{\text{SEQ}}$ is the preamble sequence duration determined by the selected \ac{PRACH} format. Unlike \ac{ZC} sequences, whose quadratic phase in $k$ couples delay and Doppler in the correlation output, the logarithmic phase law produces a nonlinear sweep that maintains this decoupling across the allocated \ac{PRACH} band.

\subsubsection{OFDM-Compatible Transmission}
To generate the \ac{PRACH} signal, the \ac{UE} maps the frequency-domain \ac{HFM} sequence $\{S[k]\}_{k=0}^{N_{\text{SEQ}}-1}$ block~\cite{9681944}. Let $\{\bar{S}[k]\}_{k=0}^{N-1}$ denote the \ac{IDFT} input, where
\begin{equation}
  \bar{S}[k]
  =
  \begin{cases}
    S[k], & 0 \leq k \leq N_{\text{SEQ}}-1, \\
    0,    & N_{\text{SEQ}} \leq k \leq N-1.
  \end{cases}
\end{equation}
Without loss of generality, the subcarrier mapping is shown starting from index zero; in a practical \ac{NR} deployment, the sequence is placed at the configured \ac{PRACH} frequency offset. The time-domain transmit sequence is obtained as
\begin{equation}
  \bar{s}[n]
  =
  \frac{1}{N}
  \sum_{k=0}^{N-1}
  \bar{S}[k]\,
  e^{\,j2\pi kn/N},
  \qquad n = 0, \dots, N-1.
\end{equation}
The transmitted preamble is constructed by prepending the \ac{CP} and appending the \ac{GP} per the \ac{PRACH} timing structure of 3GPP~TS~38.211~\cite{3gpp-ts-38.211}, identical to the standard \ac{NR} \ac{ZC} generation~\cite{9681944} with only $S[k]$ replaced. The proposed preamble is thus a constant-modulus frequency-domain sequence, not a single-carrier time-domain chirp: its logarithmic phase law spans the full
length $N_{\text{SEQ}}$ across the $L$ \ac{OFDM} symbols of the \ac{PRACH} occasion, so the Doppler-tolerant correlation is a property of the complete sequence, not of any individual symbol. A larger $L$ extends $N_{\text{SEQ}}$, giving the logarithmic phase sufficient span to develop the nonlinear curvature that distinguishes it from the quadratic \ac{ZC} law.

\subsubsection{Preamble Codebook Design}
\label{sec:codebook}

To support multiple preamble identities, a preamble-specific scaling factor $c_m$ is introduced into the frequency-domain \ac{HFM} sequence of~\eqref{tx_eq1}, yielding
\begin{equation}\label{eq:sm}
\resizebox{\columnwidth}{!}{$\displaystyle
  S_m[k] = \exp\!\left( j\,\frac{2\pi c_m}{\beta_{\text{HFM}}}
  \ln\!\left(1 + \frac{\beta_{\text{HFM}} k}{N_{\text{SEQ}}}\right)\right),\quad
  k = 0,\dots,N_{\text{SEQ}}-1
$}
\end{equation}
where $c_m$ controls the rate of phase evolution for preamble index $m$. Setting $c_m = 1$ recovers the canonical \ac{HFM}-inspired preamble; distinct values of $c_m$ produce preambles with different phase trajectories that share the same logarithmic structure and hence
retain the Doppler-tolerance property.

To guarantee low cross-correlation between any two distinct preambles, the scaling factors are chosen to satisfy
\begin{equation}
  \label{eq:ID_condition}
  \left| c_m - c_{m'} \right| \cdot \frac{2\pi}{\beta_{\text{HFM}}}\,
    \ln\!\left(1 + \beta_{\text{HFM}} f_L T_{\mathrm{SEQ}}\right) \geq 2\pi, 
    \qquad m \neq m',
\end{equation}
which ensures that the accumulated phase difference between preambles $m$ and $m'$ reaches at least $2\pi$ over the full preamble duration, preventing coherent cross-correlation buildup. The minimum spacing between consecutive scaling factors follows as
\begin{equation}
  \label{eq:delta_c}
  \Delta c
  =
  \frac{\beta_{\text{HFM}}}
       {\ln\!\left(1 + \beta_{\text{HFM}}\, f_L\, T_{\text{SEQ}}\right)}.
\end{equation}
The uniform assignment then forms the codebook
\begin{equation}
  \label{eq:codebook}
  c_m = c_0 + m\,\Delta c,
  \quad m = 0, 1, \dots, M-1,
\end{equation}
where $M$ is the desired codebook size. The initial scaling factor $c_0 > 0$ is placed so that the codebook is centered on the canonical waveform $c_m = 1$, which produces the sweep that exactly fills the allocated band, giving
\begin{equation}
  \label{eq:c0}
  c_0 = 1 - \frac{M-1}{2}\,\Delta c.
\end{equation}
Since all preambles are generated with scaling factors close to the $c_m = 1$ reference, they occupy the full allocated band and exhibit comparable main-lobe sharpness, while the spacing in~\eqref{eq:delta_c} guarantees the low cross-correlation required for reliable identification. For a given \ac{PRACH} format, $f_L$, $f_H$, and $T_{\mathrm{SEQ}}$ are fixed by the configured numerology, so $\beta_{\text{HFM}}$ is a constant shared by all preamble indices; only $c_m$ varies.

%%%%%%%%%%%%%%%%%%%%%%%%%%%%%%%%%%%%%%%%%%%%%%%%%%%%%%%%%%%%%%%%%%%%%%%%%%%%%%%%%%%%%%%
\section{Receiver-Side Processing}
\label{Sec_III}

The \ac{PRACH} receiver at the \ac{LEO}-based gNB detects active preamble transmissions and estimates the associated timing offsets from the received uplink signal. We adopt the conventional matched-filter receiver, so that the resulting performance can be attributed solely to the design of the proposed preamble. Owing to the Doppler-robust correlation structure of the proposed preamble, this standard receiver retains accurate delay estimation over a significantly coarser Doppler hypothesis grid ($\mathcal{V}_{\text{HFM}} \ll \mathcal{V}_{\text{ZC}}$), reducing overall complexity.

\subsection{Front-End Processing}
The received uplink signal is downconverted, sampled per the configured \ac{PRACH} numerology, and \ac{CP}-removed. For each \ac{OFDM} symbol, an $N$-point \ac{DFT} extracts the allocated \ac{PRACH} subcarriers while discarding out-of-band ones; the extracted blocks are concatenated into $\{\hat{Y}[k]\}_{k=0}^{N_{\text{SEQ}}-1}$ and an $N_{\text{SEQ}}$-point \ac{IDFT} reconstructs the time-domain preamble. This chain is identical to the conventional \ac{ZC}-based \ac{NR} \ac{PRACH} receiver~\cite{9681944,10571108}, and since transmitter and receiver share the same $N_{\text{SEQ}}$-point \ac{DFT}/\ac{IDFT}, the \ac{HFM} phase structure is preserved. The reconstructed signal is
\begin{equation}\label{combined_r}
  r[n]
  =
  \sum_{u \in \mathcal{U}}
  h_u\, s_{m_u}[n - d_u]\,
  e^{j2\pi \nu_u n}
  + w[n],
\end{equation}
where $s_{m_u}[n]$ is the time-domain \ac{HFM} preamble for index $m_u$, obtained via the $N_{\text{SEQ}}$-point \ac{IDFT} of $S_{m_u}[k]$, and $d_u = \tau_u / T_s$ and $\nu_u = f_{D,u} T_s$ are the discrete-time timing offset (in samples) and normalized Doppler shift of the $u$-th \ac{UE}.

\subsection{Matched Filtering and Timing Estimation}
To maintain detection sensitivity across the full Doppler uncertainty range $[-\nu_{\max},\, \nu_{\max}]$, the receiver evaluates a set of uniformly spaced Doppler hypotheses $\mathcal{V}$. For each $\nu \in \mathcal{V}$, the received signal
is frequency de-rotated as $ r_{\nu}[n] = r[n]\, e^{-j2\pi \nu n}$. A residual offset of up to $\pm\Delta\nu_{\text{bin}}/2$ remains within each bin. For \ac{ZC} sequences, this residual displaces the correlation peak, necessitating a fine Doppler grid~\cite{9681944}. The \ac{HFM} correlation peak, however,
remains at the true delay irrespective of the residual offset, permitting a coarser grid with $|\mathcal{V}_{\text{HFM}}| \ll |\mathcal{V}_{\text{ZC}}|$. For each hypothesis $(m, \nu)$, the matched-filter output is computed on the de-rotated signal $r_\nu[n]$ as
\begin{equation}\label{eq:mf}
  \Lambda_m(d, \nu)
  =
  \sum_{n=0}^{N_{\text{SEQ}}-1}
  r_\nu[n]\,
  s_m^{*}[n - d],
  \qquad d \in \mathcal{D}.
\end{equation}
The magnitude of the matched-filter output,
\begin{equation}
  z_m(d, \nu) = \left|\Lambda_m(d, \nu)\right|,
\end{equation}
constitutes the test statistic for preamble detection and timing estimation.
To elucidate the role of the \ac{HFM}-inspired preamble in decoupling timing from Doppler, consider a single active \ac{UE} transmitting preamble $m_0$ with timing offset $d_0$ and Doppler shift $\nu_0$. Substituting into~\eqref{eq:mf} and separating the signal and noise contributions yields
\begin{equation}\label{eq:mf_expanded}
\resizebox{\columnwidth}{!}{$\displaystyle
  \Lambda_m(d, \nu)
  = h\, e^{-j2\pi (\nu - \nu_0) d_0}\,
    \chi_{m,m_0}(d - d_0,\, \nu - \nu_0) + \tilde{w}_m(d, \nu)
$}
\end{equation}
where
\begin{equation}\label{AF_def}
  \chi_{m,m_0}(\Delta d,\, \Delta\nu)
  \triangleq
  \sum_{n=0}^{N_{\text{SEQ}}-1}
  s_{m_0}[n]\, s_m^{*}[n - \Delta d]\, e^{-j2\pi \Delta\nu\, n}.
\end{equation}
denotes the delay--Doppler cross-\ac{AF} between the transmitted preamble $m_0$ and the receiver hypothesis $m$, and $\tilde{w}_m(d, \nu)$ is the filtered noise term. For the correct preamble hypothesis $m = m_0$, \eqref{AF_def} reduces to the self-\ac{AF} $\chi_{m_0,m_0}(\Delta d, \Delta\nu)$, which governs the sharpness and localization of the correlation peak.

\subsection{Hypothesis Testing and Detection Rule}

At each hypothesis point $(m, d, \nu)$ in the preamble--delay--Doppler domain, preamble detection is formulated as a binary hypothesis test,
\begin{equation}
\begin{aligned}
  \mathcal{H}_0 &:\quad z_m(d, \nu) = |\tilde{w}_m(d, \nu)|, \\
  \mathcal{H}_1 &:\quad z_m(d, \nu)
    = |h\,\chi_{m,m_0}(d - d_0,\, \nu - \nu_0)
      + \tilde{w}_m(d, \nu)|,
\end{aligned}
\end{equation}
where $\mathcal{H}_0$ corresponds to the absence of any preamble transmission at the evaluated hypothesis, so that the matched-filter output contains only filtered noise, and
$\mathcal{H}_1$ corresponds to the presence of a transmitted preamble $m_0$ with true delay $d_0$ and Doppler offset $\nu_0$. Under $\mathcal{H}_0$, the matched-filter output $\Lambda_m(d,\nu)$ is a circularly symmetric complex Gaussian random variable, and
$z_m(d,\nu) = |\Lambda_m(d,\nu)|$ follows a Rayleigh distribution. A preamble transmission is declared present whenever
\begin{equation}
  z_m(d, \nu) > \eta,
\end{equation}
where the detection threshold $\eta$ is selected to satisfy a prescribed false-alarm probability,
\begin{equation}
  P_{\text{FA}}
  \triangleq
  \Pr\!\left(z_m(d, \nu) > \eta \mid \mathcal{H}_0\right).
\end{equation}
The explicit relationship between $\eta$ and $P_{\text{FA}}$ is derived in Section~\ref{Sec:performance_analysis}. Given a detection, the joint timing and Doppler estimates for preamble $m$ are obtained as
\begin{equation}
  (\hat{d}_m, \hat{\nu}_m)
  =
  \arg\max_{d \in \mathcal{D},\, \nu \in \mathcal{V}}\,
  z_m(d, \nu),
\end{equation}
and the detected preamble index is identified as the $m$ for which $z_m(\hat{d}_m, \hat{\nu}_m)$ exceeds $\eta$.

\subsection{Multi-User Scenario}

In practical random access, multiple \acp{UE} may transmit different preambles within the same \ac{PRACH} occasion. In this case, the matched-filter output for preamble hypothesis $m$ admits the decomposition~\cite{9340403}
\begin{equation}\label{multi_ID}
\begin{aligned}
  \Lambda_m(d, \nu)
  &=
  \sum_{u \in \mathcal{U}}
  h_u\, e^{-j2\pi (\nu - \nu_u) d_u}\,
  \chi_{m,m_u}(d - d_u,\, \nu - \nu_u) \\
  &\quad + \tilde{w}_m(d, \nu),
\end{aligned}
\end{equation}
which reveals that multi-user interference manifests as cross-\ac{AF} terms $\chi_{m,m_u}$ evaluated at the delay and Doppler offsets of interfering \acp{UE}. For the target preamble
$m = m_u$, the corresponding self-\ac{AF} term produces a detectable peak at $d = d_u$, while the cross-\ac{AF} terms from other active preambles contribute to structured interference. The codebook design condition~\eqref{eq:ID_condition} ensures that the accumulated phase difference between any two distinct preambles exceeds $2\pi$, suppressing the cross-\ac{AF} sidelobes and enabling reliable separation of concurrent preamble transmissions in the delay domain.

The final receiver output is the set of detected preamble indices and their associated timing and Doppler estimates,
\begin{equation}
  \{(\hat{m}_u,\, \hat{d}_u,\, \hat{\nu}_u)\}_{u \in
  \hat{\mathcal{U}}},
\end{equation}
where $\hat{\mathcal{U}}$ denotes the inferred set of active \acp{UE}. The complete receiver processing is summarized in Algorithm~\ref{alg:receiver}. The explicit appearance of the \ac{AF} throughout the receiver formulation provides a direct analytical link to the performance analysis in Section~\ref{Sec:performance_analysis}, where the delay--Doppler ambiguity structure of the \ac{ZC} and \ac{HFM}-inspired preambles is characterized and connected to detection probability, timing accuracy, and Doppler resilience.

\begin{algorithm}[t]
\caption{HFM-Based PRACH Detection and Timing Estimation}
\label{alg:receiver}
\begin{algorithmic}
\REQUIRE Received signal $r[n]$ after front-end processing,
         preamble codebook $\{s_m[n]\}_{m=0}^{M-1}$,
         delay search set $\mathcal{D}$,
         Doppler hypothesis set $\mathcal{V}$,
         detection threshold $\eta$
\ENSURE  Detected set
         $\{(\hat{m}_u,\, \hat{d}_u,\, \hat{\nu}_u)\}_{u \in
         \hat{\mathcal{U}}}$

\STATE \textbf{Initialize:} $\hat{\mathcal{U}} \leftarrow \emptyset$

\FOR{each Doppler hypothesis $\nu \in \mathcal{V}$}
  \STATE $r_{\nu}[n] \leftarrow r[n]\,
         e^{-j2\pi \nu n T_s}$
\ENDFOR

\FOR{each preamble index $m = 0, \dots, M-1$}
  \FOR{each Doppler hypothesis $\nu \in \mathcal{V}$}
    \FOR{each delay hypothesis $d \in \mathcal{D}$}
      \STATE $\Lambda_m(d, \nu) \leftarrow
             \sum_{n=0}^{N_{\text{SEQ}}-1}
             r_{\nu}[n]\, s_m^{*}[n - d]$
      \STATE $z_m(d, \nu) \leftarrow
             |\Lambda_m(d, \nu)|$
    \ENDFOR
  \ENDFOR

  \STATE $(\hat{d}_m, \hat{\nu}_m) \leftarrow
         \arg\max_{d,\nu}\, z_m(d, \nu)$

  \IF{$z_m(\hat{d}_m, \hat{\nu}_m) > \eta$}
    \STATE $\hat{\mathcal{U}} \leftarrow
           \hat{\mathcal{U}} \cup
           \{(m,\, \hat{d}_m,\, \hat{\nu}_m)\}$
  \ENDIF
\ENDFOR

\RETURN $\{(\hat{m}_u,\, \hat{d}_u,\, \hat{\nu}_u)\}_{u \in
        \hat{\mathcal{U}}}$
\end{algorithmic}
\end{algorithm}

%%%%%%%%%%%%%%%%%%%%%%%%%%%%%

\section{Performance Analysis}
\label{Sec:performance_analysis}
This section analyzes the detection and identification performance of the proposed \ac{HFM} preamble and baseline approaches, proceeding from self-\ac{AF} characterization under Doppler, through cross-\ac{AF} based preamble separability, to false-alarm and detection probability analysis. The computational complexity of the proposed scheme is also analyzed.

\subsection{Self-Ambiguity Function Analysis}
\label{sec:ambiguity_analysis}
From the receiver formulation in Section~\ref{Sec_III}, the matched-filter output under a single active \ac{UE} is governed by the self-\ac{AF} $\chi_{m_0,m_0}(\Delta d, \Delta\nu)$ defined in~\eqref{AF_def}. In the absence of noise, timing estimation reduces to maximizing $|\chi_{m_0,m_0}(d - d_0, \nu - \nu_0)|$ over the delay search set $\mathcal{D}$, so robustness under Doppler is determined by how the self-\ac{AF} peak moves along the delay axis with the frequency offset $\Delta\nu$.

\subsubsection{Zadoff--Chu Preamble}

For an odd sequence length $N$ and root index $y$ coprime with $N$,
the \ac{ZC} preamble is~\cite{10001203}
\begin{equation}\label{ZC_def}
  s_{\text{ZC}}[n]
  =
  \exp\!\left(-\frac{j\pi y}{N}\,n(n+1)\right).
\end{equation}

\begin{proposition}[ZC Delay--Doppler Coupling]
\label{prop:zc}
The self-\ac{AF} of the \ac{ZC} sequence~\eqref{ZC_def} satisfies
\begin{equation}\label{eq:ZC_AF}
  |\chi_{\text{ZC}}(d, \nu)|
  =
  \left|
  \frac{\sin\!\Big(\pi N\big(\nu + \frac{y}{N}d\big)\Big)}
       {\sin\!\Big(\pi\big(\nu + \frac{y}{N}d\big)\Big)}
  \right|,
\end{equation}
which attains its maxima when
\begin{equation}\label{eq:ZC_peak}
  \nu + \frac{y}{N}\,d = p, \qquad p \in \mathbb{Z}.
\end{equation}
\end{proposition}

Substituting~\eqref{ZC_def} into~\eqref{AF_def} and completing the square reduces the resulting geometric series to the Dirichlet kernel~\eqref{eq:ZC_AF}, with the peak condition~\eqref{eq:ZC_peak} obtained by setting the kernel argument to an integer. For nonzero $\nu$, the delay maxima $\mathcal{D}_{\text{peak}} = \{d \mid \nu + (y/N)d \in \mathbb{Z}\}$ form multiple peaks of comparable magnitude spaced $N/y$ samples apart, creating a timing ambiguity that worsens as $\nu$ grows.

\subsubsection{LFM Preamble}

The \ac{LFM} preamble with chirp rate $\beta_{\text{LFM}}$ is defined as
\begin{equation}\label{LFM_def}
  s_{\text{LFM}}[n]
  =
  \exp\!\left(j\pi\beta_{\text{LFM}}\, n^2\right),
  \quad \quad n = 0, \dots, N-1,
\end{equation}
where $\beta_{\text{LFM}} = (f_H - f_L)\,T_s / N$.

\begin{proposition}[LFM Timing Bias]
\label{prop:lfm}
The self-\ac{AF} of the \ac{LFM} sequence~\eqref{LFM_def}
satisfies
\begin{equation}\label{eq:LFM_AF}
  |\chi_{\text{LFM}}(d, \nu)|
  =
  \left|
  \frac{\sin\!\big(\pi N(\beta_{\text{LFM}} d - \nu)\big)}
       {\sin\!\big(\pi(\beta_{\text{LFM}} d - \nu)\big)}
  \right|,
\end{equation}
which is maximized along the ridge $\mathcal{R} = \{(d, \nu) \mid \beta_{\text{LFM}} d = \nu\}$. For a true delay $d_0$ and Doppler offset $\nu_0$, the delay estimate is $\hat{d} = d_0 + \frac{\nu_0}{\beta}$, exhibiting a deterministic timing bias proportional to the Doppler offset.
\end{proposition}

Substituting~\eqref{LFM_def} into~\eqref{AF_def} and factoring the quadratic phase yields a geometric series in $\beta_{\text{LFM}} d - \nu$, giving the Dirichlet kernel~\eqref{eq:LFM_AF}, whose peak at $\beta_{\text{LFM}} d - \nu = 0$ produces the bias. This bias is a structural consequence of the affine delay--Doppler coupling and persists even in the absence of noise.

\subsubsection{HFM-inspired Preamble}

The proposed \ac{HFM} frequency-domain preamble $S_m[k]$ defined in~\eqref{eq:sm} produces, after $N_{\text{SEQ}}$-point \ac{IDFT}, the time-domain sequence $s_m[n]$ used for matched filtering at the receiver. Applying the self-\ac{AF} definition~\eqref{AF_def} yields
\begin{equation}\label{eq:HFM_AF}
\resizebox{\columnwidth}{!}{$\displaystyle
  \chi_{\text{HFM}}(d, \nu)
  = \sum_{n=0}^{N_{\text{SEQ}}-1}
    \exp\!\left(
      j\,\frac{2\pi c_m}{\beta_{\text{HFM}}}
      \ln\frac{1 + \frac{\beta_{\text{HFM}}\, n}{N_{\text{SEQ}}}}
              {1 + \frac{\beta_{\text{HFM}}(n-d)}{N_{\text{SEQ}}}}
    \right)
    e^{-j2\pi\nu n}
$}
\end{equation}

\begin{theorem}[HFM Timing Localization Without Doppler Bias]
\label{thm:hfm}
The self-\ac{AF} of the \ac{HFM}-inspired preamble in~\eqref{eq:HFM_AF} does not admit an affine delay--Doppler peak condition of the form $\alpha d + \gamma\nu = k$ for any constants $\alpha$, $\gamma$. Consequently, a Doppler offset $\nu$ induces neither a deterministic displacement of the delay maximizer (as in \ac{LFM}) nor a periodic lattice of delay maxima (as in \ac{ZC}). At $d = 0$ the Doppler contributes only a delay-independent linear phase ramp, so it attenuates the peak according to a Dirichlet envelope rather than shifting it; for $|\nu| < 1/N_{\text{SEQ}}$ the delay maximizer remains at $d = 0$.
\end{theorem}

The \ac{AF} in~\eqref{eq:HFM_AF} is a coherent sum of unit-magnitude phasors,
\begin{equation}\label{eq:phasor}
  \chi_{\text{HFM}}(d, \nu)
  =
  \sum_{n=0}^{N_{\text{SEQ}}-1}
  e^{j\psi_n(d,\nu)},
\end{equation}
with phase
\begin{equation}
  \psi_n(d, \nu)
  =
  \frac{2\pi c_m}{\beta}
  \ln\frac{1 + \frac{\beta\, n}{N_{\text{SEQ}}}}
          {1 + \frac{\beta(n-d)}{N_{\text{SEQ}}}}
  - 2\pi\nu n.
\end{equation}
Doppler enters~\eqref{eq:phasor} only through the delay-independent ramp $-2\pi\nu n$. At $d=0$ the logarithmic term vanishes for all $n$, and~\eqref{eq:phasor} reduces to $\sum_{n} e^{-j2\pi\nu n}$, of magnitude $|\chi_{\text{HFM}}(0,\nu)| = |\sin(\pi N_{\text{SEQ}}\nu)/\sin(\pi\nu)|$. This Dirichlet envelope attenuates with $|\nu|$ and nulls at $\nu = p/N_{\text{SEQ}}$, $p \in \mathbb{Z}\setminus\{0\}$; being delay-independent, it lowers the peak without displacing it. For $d \neq 0$ the logarithmic term is nonlinear in $n$, so the phase cannot be written as a ramp $2\pi(\alpha d + \gamma\nu)n$; the maxima thus admit no affine locus $\alpha d + \gamma\nu = k$, ruling out the \ac{LFM} ridge and \ac{ZC} lattice. Over $|\nu| < 1/N_{\text{SEQ}}$ the maximizer stays at $d=0$.

\begin{corollary}[Timing Estimation Without Doppler Bias]
\label{cor:unbiased}
For a received preamble with true delay $d_0$ and Doppler offset $\nu_0$, let $\nu_q \in V$ be the nearest hypothesis and $\Delta\nu = \nu_0 - \nu_q$ the residual after de-rotation. In the absence of noise, the \ac{HFM} timing estimate $\hat{d} = \arg\max_{d}\, |\chi_{\text{HFM}}(d - d_0, \Delta\nu)|$ satisfies $\hat{d} = d_0$ for $|\Delta\nu| < 1/N_{\text{SEQ}}$; that is, it carries no deterministic Doppler-dependent bias. This contrasts with the \ac{ZC} and \ac{LFM} estimators, which exhibit the Doppler-dependent biases established in Propositions~\ref{prop:zc} and~\ref{prop:lfm}.
\end{corollary}

By Theorem~\ref{thm:hfm}, the self-\ac{AF} delay response stays centered at $\Delta d = 0$ for residual offsets $|\Delta\nu| < 1/N_{\text{SEQ}}$, so the maximizer satisfies $\hat{d} = d_0$ with no Doppler-dependent shift. The peak is attenuated by the Dirichlet envelope but not displaced; the grid $\mathcal{V}$ bounds $|\Delta\nu|$ within a bin, restoring the coherent gain.

\subsection{Cross-Ambiguity Analysis and Preamble Identification}
\label{subsec:cross_af}

In multi-user random access, the receiver must distinguish the transmitted preamble $m_0$ from all other candidate indices $m \neq m_0$. From~\eqref{multi_ID}, incorrect preamble hypotheses contribute cross-\ac{AF} terms $\chi_{m,m_0}(d - d_u, \nu - \nu_u)$ to the matched-filter output. Reliable preamble identification requires that these cross-\ac{AF} terms remain sufficiently below the self-\ac{AF} peak across the entire delay--Doppler search region. Preamble misidentification occurs when a cross-\ac{AF} sidelobe exceeds the self-\ac{AF} peak, i.e.,
\begin{equation}\label{eq:misid}
  \max_{m \neq m_0}\,
  \max_{d, \nu}\,
  |\chi_{m,m_0}(d - d_0, \nu - \nu_0)|
  \;\geq\;
  |\chi_{m_0,m_0}(0, 0)|.
\end{equation}
Two standard metrics quantify the susceptibility of a given waveform to this condition. The \ac{PSLR} measures the ratio between the largest cross-\ac{AF} magnitude and the self-\ac{AF} peak,
\begin{equation}\label{eq:pslr}
  \text{PSLR}
  =
  20\log_{10}\!\left(
    \frac{\displaystyle
      \max_{m \neq m_0}\,
      \max_{d, \nu}\,
      |\chi_{m,m_0}(d, \nu)|}
         {|\chi_{m_0,m_0}(0, 0)|}
  \right),
\end{equation}
while the \ac{ISLR} measures the total cross-\ac{AF} energy normalized by the mainlobe energy,
\begin{equation}\label{eq:islr}
  \text{ISLR}
  =
  10\log_{10}\!\left(
    \frac{\displaystyle
      \sum_{m \neq m_0}
      \sum_{d, \nu}
      |\chi_{m,m_0}(d, \nu)|^2}
         {|\chi_{m_0,m_0}(0, 0)|^2}
  \right).
\end{equation}
Negative \ac{PSLR} values indicate that all cross-\ac{AF} sidelobes are strictly below the mainlobe, ensuring reliable preamble identification. Lower \ac{ISLR} values indicate that the total interference energy from incorrect preamble hypotheses is well suppressed relative to the desired signal.

\subsubsection{ZC Cross-Ambiguity}

For \ac{ZC} preambles with distinct roots $u_m \neq u_{m_0}$, the cross-\ac{AF} follows from~\eqref{AF_def} as
\begin{equation}\label{eq:zc_cross}
  |\chi_{m,m_0}^{\text{ZC}}(d, \nu)|
  =
  \left|
  \frac{\sin\!\Big(\pi N\big(\nu +
    \frac{u_m - u_{m_0}}{N}d\big)\Big)}
       {\sin\!\Big(\pi\big(\nu +
    \frac{u_m - u_{m_0}}{N}d\big)\Big)}
  \right|,
\end{equation}
which attains maxima equal to $N$ whenever
\begin{equation}\label{eq:zc_cross_peak}
  \nu + \frac{u_m - u_{m_0}}{N}\,d = p,
  \qquad p \in \mathbb{Z}.
\end{equation}
Since the self-\ac{AF} peak also equals $N$, these maxima are indistinguishable from the mainlobe, forming a periodic lattice of equally strong incorrect-hypothesis peaks. The \ac{ZC} cyclic autocorrelation suppresses these terms at zero delay only under negligible Doppler; for large $\nu$ in \ac{NTN}, the lattice points shift into the active delay search window, so multiple incorrect hypotheses exceed the threshold and \ac{ZC}-based identification becomes unreliable.

\subsubsection{LFM Cross-Ambiguity}

For \ac{LFM} preambles with distinct chirp rates
$\beta_{{\text{LFM}},m} \neq \beta_{{\text{LFM}},m_0}$, the cross-\ac{AF} magnitude is
\begin{equation}\label{eq:lfm_cross}
  |\chi_{m,m_0}^{\text{LFM}}(d, \nu)|
  =
  \left|
  \frac{\sin\!\big(\pi N((\beta_{{\text{LFM}},m} - \beta_{{\text{LFM}},m_0})d - \nu)\big)}
       {\sin\!\big(\pi((\beta_{{\text{LFM}},m} - \beta_{{\text{LFM}},m_0})d - \nu)\big)}
  \right|,
\end{equation}
which is maximized along the affine set
\begin{equation}\label{eq:lfm_cross_ridge}
  \mathcal{R}^{\text{cross}}
  =
  \big\{(d, \nu) \mid (\beta_{{\text{LFM}},m} - \beta_{{\text{LFM}},m_0})d = \nu\big\}.
\end{equation}
Along $\mathcal{R}^{\text{cross}}$ the cross-\ac{AF} equals the self-\ac{AF} peak, so every ridge point is a potential misidentification, and the total cross-\ac{AF} energy grows with the search region. As Doppler uncertainty increases in \ac{NTN}, a larger portion of the ridge enters the search window, elevating the \ac{ISLR} and the probability that an incorrect preamble crosses the threshold.

\begin{remark}
Although \ac{LFM} offers improved Doppler tolerance for timing estimation over \ac{ZC} (Proposition~\ref{prop:lfm}), its cross-\ac{AF} ridge lets incorrect hypotheses produce peaks as strong as the correct preamble, making \ac{LFM} unsuitable for multi-user random access requiring reliable identification.
\end{remark}

\subsubsection{HFM Cross-Ambiguity}

For the proposed \ac{HFM} preambles with distinct scaling factors $c_m \neq c_{m_0}$, the cross-\ac{AF} follows
from~\eqref{AF_def} as
\begin{equation}\label{eq:hfm_cross}
  \chi_{m,m_0}^{\text{HFM}}(d, \nu)
  =
  \sum_{n=0}^{N_{\text{SEQ}}-1}
  \exp\!\big(j\psi_{m,m_0}(n; d, \nu)\big),
\end{equation}
where
\begin{equation}\label{eq:hfm_cross_phase}
\resizebox{\columnwidth}{!}{$\displaystyle
  \psi_{m,m_0}(n; d, \nu)
  = \frac{2\pi(c_{m_0} - c_m)}{\beta_{\text{HFM}}}
    \ln\frac{1 + \frac{\beta_{\text{HFM}}\, n}{N_{\text{SEQ}}}}
            {1 + \frac{\beta_{\text{HFM}}(n-d)}{N_{\text{SEQ}}}} - 2\pi\nu n
$}
\end{equation}
The phase in~\eqref{eq:hfm_cross_phase} is nonlinear in $n$ through the logarithmic ratio, with the nonlinearity scaled by $(c_{m_0} - c_m)$. This differs fundamentally from the \ac{ZC} and \ac{LFM} cross-\ac{AF} expressions, which reduce to Dirichlet kernels in affine combinations of $(d, \nu)$.

\begin{proposition}[HFM Preamble Separability]
\label{prop:hfm_cross}
Under the codebook design condition~\eqref{eq:ID_condition}, the
cross-\ac{AF} of the \ac{HFM}-inspired preamble satisfies
\begin{equation}\label{eq:hfm_sep}
  \max_{d, \nu}\,
  |\chi_{m,m_0}^{\text{HFM}}(d, \nu)|
  \;<\;
  |\chi_{m_0,m_0}^{\text{HFM}}(0, 0)|,
  \qquad \forall\, m \neq m_0.
\end{equation}
\end{proposition}

Coherent accumulation of~\eqref{eq:hfm_cross} requires the per-sample phase increments $\Delta\psi_n = \psi_{m,m_0}(n; d, \nu) - \psi_{m,m_0}(n-1; d, \nu)$ to have low dispersion across $n$. The codebook condition~\eqref{eq:ID_condition} forces the total phase excursion over the summation support to span at least $2\pi$ for any $m \neq m_0$, so the phasors in~\eqref{eq:hfm_cross} complete at least one full rotation and cannot align at any $(d, \nu)$. Equivalently, the triangle-inequality bound $|\chi_{m,m_0}^{\text{HFM}}(d, \nu)| \leq \sum_{n=0}^{N_{\text{SEQ}}-1} |e^{j\psi_{m,m_0}(n; d, \nu)}| = N_{\text{SEQ}}$ is attained only under perfect alignment, which~\eqref{eq:ID_condition} precludes; the resulting destructive interference keeps the cross-\ac{AF} strictly below the self-\ac{AF} peak $N_{\text{SEQ}}$ at $(d, \nu) = (0, 0)$, and distributes the residual sidelobe energy diffusely rather than along any deterministic trajectory, establishing~\eqref{eq:hfm_sep}.

\begin{remark}
The separability in Proposition~\ref{prop:hfm_cross} follows from the nonlinear \ac{HFM} phase structure together with the minimum spacing on $\{c_m\}$. Unlike the \ac{ZC} and \ac{LFM} cases, where separation degrades as the lattice and ridge enter the search window with increasing Doppler, the \ac{HFM} cross-\ac{AF} suppression holds uniformly over the delay--Doppler plane, making the codebook inherently suited to \ac{NTN} occasions where large Doppler uncertainty and multi-user access coexist.
\end{remark}

Table~\ref{tab:af_comparison} summarizes the delay--Doppler ambiguity properties of the three waveforms.

\begin{table}[t]
\centering
\caption{Comparison of Delay--Doppler Ambiguity Properties. Timing bias is expressed in samples, with $\nu$ the normalized Doppler offset.}
\label{tab:af_comparison}
\begin{tabular}{lccc}
\hline
\textbf{Property} & \textbf{ZC} & \textbf{LFM} & \textbf{HFM} \\
\hline
Timing bias under Doppler
  & $y^{-1} N\, \nu$ & $\nu/\beta_{\text{LFM}}$ & $0^{\dagger}$ \\
Self-AF peak shift
  & Yes & Yes & No$^{\ddagger}$ \\
Cross-AF structure
  & Lattice & Ridge & Diffuse \\
Preamble separability
  & Degraded & Degraded & Preserved \\
\hline
\end{tabular}

\vspace{2pt}
\begin{flushleft}
\footnotesize
$^{\dagger}$ No deterministic Doppler-dependent timing bias; by Theorem~\ref{thm:hfm} the peak does not migrate within a Doppler bin. \\
$^{\ddagger}$ The peak attenuates with $\nu$ (Dirichlet envelope) but is not displaced.
\end{flushleft}
\end{table}

\subsection{False Alarm and Detection Probability}
\label{subsec:pfa_pd}

The detection rule formulated in Section~\ref{Sec_III} declares a preamble present whenever the test statistic $z_m(d, \nu)$ exceeds a threshold $\eta$ at some hypothesis point $(m, d, \nu)$. This subsection derives the false alarm and detection probabilities that
govern the receiver operating characteristics.

\subsubsection{False Alarm Probability}

Under $\mathcal{H}_0$, the matched-filter output $\Lambda_m(d,\nu)$ contains only filtered noise and is circularly symmetric complex Gaussian with variance $\sigma_\Lambda^2$. The test statistic $z_m(d, \nu) = |\Lambda_m(d, \nu)|$ therefore follows a Rayleigh distribution,
\begin{equation}\label{eq:rayleigh_pdf}
  p_Z(z)
  =
  \frac{2z}{\sigma_\Lambda^2}
  \exp\!\left(-\frac{z^2}{\sigma_\Lambda^2}\right),
  \qquad z \geq 0.
\end{equation}
The probability that the test statistic exceeds the threshold
$\eta$ in the absence of any transmitted preamble is
\begin{equation}\label{eq:pfa}
  P_{\text{FA}}
  =
  \Pr(z_m(d, \nu) > \eta \mid \mathcal{H}_0)
  =
  \exp\!\left(-\frac{\eta^2}{\sigma_\Lambda^2}\right).
\end{equation}
Inverting~\eqref{eq:pfa} yields the detection threshold, $\eta= \sigma_\Lambda\sqrt{-\ln P_{\text{FA}}}$ as a function of the prescribed false alarm probability~\cite{10511361}.
It establishes that the threshold is determined solely by the noise variance and the target $P_{\text{FA}}$, independently of the preamble waveform. This
common threshold applies identically to \ac{ZC}, \ac{LFM}, and
\ac{HFM} receivers.

\subsubsection{Detection Probability}
Under $\mathcal{H}_1$, the matched-filter output at the correct hypothesis $(m_0, d_0, \nu_0)$ contains a deterministic signal component proportional to the self-\ac{AF} peak $\chi_{m_0,m_0}(0, 0)$, superimposed with Gaussian noise. The instantaneous effective post-correlation \ac{SNR} is defined as
\begin{equation}\label{eq:snr}
  \gamma
  \triangleq
  \frac{|h|^2\,|\chi_{m_0,m_0}(0, 0)|^2}{\sigma_\Lambda^2},
\end{equation}
and, conditioned on the channel realization $h$, the test statistic $z_{m_0}(d_0, \nu_0)$ follows a Rician distribution with density
\begin{equation}\label{eq:rician}
\begin{aligned}
  p_Z(z)
  &=
  \frac{2z}{\sigma_\Lambda^2}
  \exp\!\left(
    -\frac{z^2 + |h\,\chi_{m_0,m_0}(0,0)|^2}{\sigma_\Lambda^2}
  \right) \\
  &\quad \times\,
  I_0\!\left(
    \frac{2z\,|h\,\chi_{m_0,m_0}(0,0)|}{\sigma_\Lambda^2}
  \right),
  \qquad z \geq 0,
\end{aligned}
\end{equation}
where $I_0(\cdot)$ is the zeroth-order modified Bessel function of the first kind. Conditioned on $h$, the detection probability follows as~\cite{10720128} $P_{\text{D}}(\gamma) = \Pr(z_{m_0}(d_0, \nu_0) > \eta \mid \mathcal{H}_1, h) = Q_1(\sqrt{2\gamma},\,\sqrt{2\eta^2/\sigma_\Lambda^2})$, where $Q_1(\cdot, \cdot)$ denotes the first-order Marcum $Q$-function. Substituting the threshold and averaging the conditional probability over the channel fading yields the average detection probability
\begin{equation}\label{eq:pd_avg}
\begin{aligned}
  P_{\text{D}}
  &=
  \mathbb{E}_h\!\left[
    Q_1\!\left(\sqrt{2\gamma},\;\sqrt{-2\ln P_{\text{FA}}}\right)
  \right] \\
  &=
  \int_0^\infty
    Q_1\!\left(\sqrt{2\gamma},\;\sqrt{-2\ln P_{\text{FA}}}\right)
    f_\gamma(\gamma)\,\mathrm{d}\gamma.
\end{aligned}
\end{equation}
where $f_\gamma(\gamma)$ is the effective-\ac{SNR} density of the narrowband Rician channel with factor $K$ and mean \ac{SNR} $\bar\gamma \triangleq \mathbb{E}[\gamma]$. Throughout, $P_{\text{D}}$ denotes this average detection probability, consistent with the Monte-Carlo curves in Section~\ref{sec:results}.
\begin{remark}
Since $|\chi_{m_0,m_0}(0,0)| = N_{\text{SEQ}}$ for all three waveforms, the effective \ac{SNR} $\gamma$ in~\eqref{eq:snr} and its density $f_\gamma$ are identical for \ac{ZC}, \ac{LFM}, and \ac{HFM}, so the average detection probability~\eqref{eq:pd_avg} is the same under perfect Doppler compensation. The waveforms differ only at incorrect hypotheses, where the interference \ac{SNR} is $\gamma_m(d,\nu) = \rho_m(d,\nu)\,\gamma$ with $\rho_m(d,\nu) \triangleq |\chi_{m,m_0}(d-d_0, \nu-\nu_0)|^2 / |\chi_{m_0,m_0}(0,0)|^2$. The \ac{ZC} and \ac{LFM} lattice and ridge structures of Section~\ref{subsec:cross_af} yield $\rho_m \approx 1$, raising the misidentification risk, whereas the codebook condition~\eqref{eq:ID_condition} enforces $\rho_m < 1$ for all $m \neq m_0$, preserving mainlobe--sidelobe separation.
\end{remark}

\begin{table}[t]
\centering
\caption{Simulation Parameters}
\label{tab:sim_params}
\begin{tabular}{lcc}
\hline
\textbf{Parameter} & \textbf{Symbol} & \textbf{Value} \\
\hline
Carrier frequency & $f_c$ & 30 GHz (Ka Band) \\
Satellite altitude & $a$ & 600 km \\
LEO velocity & $v$ & 7.5 km/s \\
Subcarrier spacing & $\Delta f$ & 120 kHz \\
PRACH preamble format & -- & Format A3 \\
Per-symbol sequence length & $L_{\text{RA}}$ & 139 \\
Number of OFDM symbols & $L$ & 6 \\
Preamble length & $N_{\text{SEQ}}$ & 834 ($= 139 \times 6$) \\
Sampling frequency & $F_s$ & 139 $\times \Delta f$ \\
False alarm probability & $P_{\text{FA}}$ & $10^{-3}$ \\
Channel model & -- & Narrowband  Rician\\
Rician $K$-factor & $K$ & 10.224 dB \\
\hline
\end{tabular}
\end{table}

\subsection{Computational Complexity}
\label{subsec:compexity}

The proposed \ac{HFM} receiver has total complexity $\mathcal{O}(M \cdot |\mathcal{V}_{\text{HFM}}| \cdot |\mathcal{D}| \cdot N_{\text{SEQ}})$, one matched filter per preamble per Doppler hypothesis. The \ac{ZC} receiver shares this structure but demands a denser grid ($|\mathcal{V}_{\text{ZC}}| \gg |\mathcal{V}_{\text{HFM}}|$) to avoid timing bias, raising its cost proportionally. The \ac{FDS-LFM} receiver~\cite{10892192} removes the Doppler search but replaces it with $M_{\text{sub}}$ subsequence correlations per preamble, each of length $\beta_{\text{FDS}} N$ ($\beta_{\text{FDS}}$ being the frequency-shifting parameter), followed by delay-shifted non-coherent combining, yielding $\mathcal{O}(M \cdot M_{\text{sub}} \cdot |\mathcal{D}| \cdot \beta_{\text{FDS}} N)$; the $M_{\text{sub}}$-fold repetition and combining overhead make it costlier than the single-correlation \ac{HFM} design. On the transmitter side, \ac{HFM} costs $\mathcal{O}(N_{\text{SEQ}})$ for one frequency-domain sequence, whereas \ac{FDS-LFM} superposes $M_{\text{sub}}$ phase-rotated subsequences before the \ac{IDFT}, incurring $\mathcal{O}(M_{\text{sub}} \cdot N_{\text{SEQ}})$.

%%%%%%%%%%%%%%%%%%%%%%%%%%%%%%%%%%%%%%%%%%

%%%%%%%%%%%%%%%%%%%%%%%%%%%%%
%%%%%%%%%%%%%%%%%%%%%%%%%%%%%%%%%%%%%%%%%%
\begin{figure*}
  \centering
  \subfloat[\label{Fig:ZC_AF_3D}]{
    \includegraphics[width=0.32\linewidth]{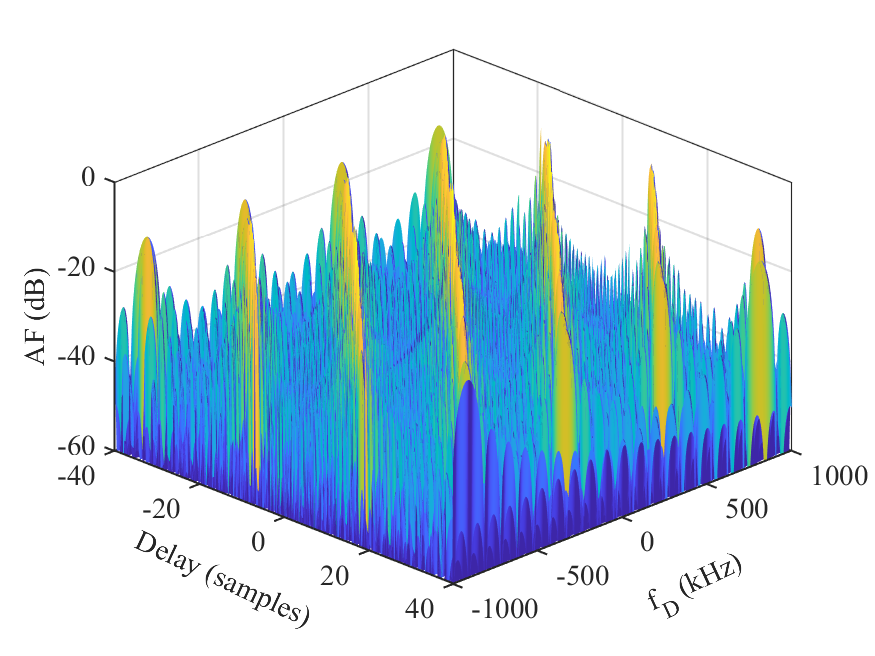}}
  \hfill
  \subfloat[\label{Fig:LFM_AF_3D}]{
    \includegraphics[width=0.32\linewidth]{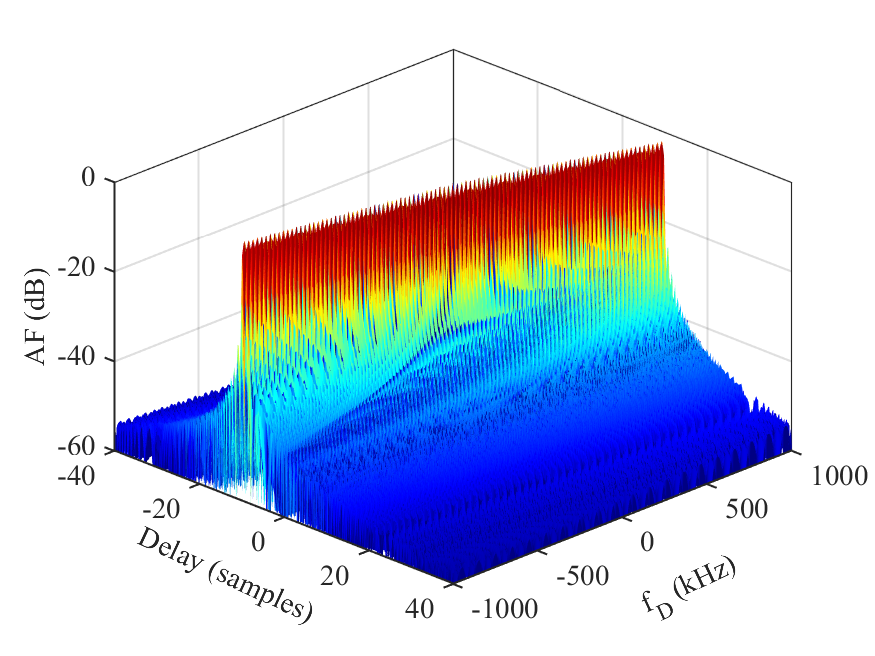}}
  \hfill
  \subfloat[\label{Fig:HFM_AF_3D}]{
    \includegraphics[width=0.32\linewidth]{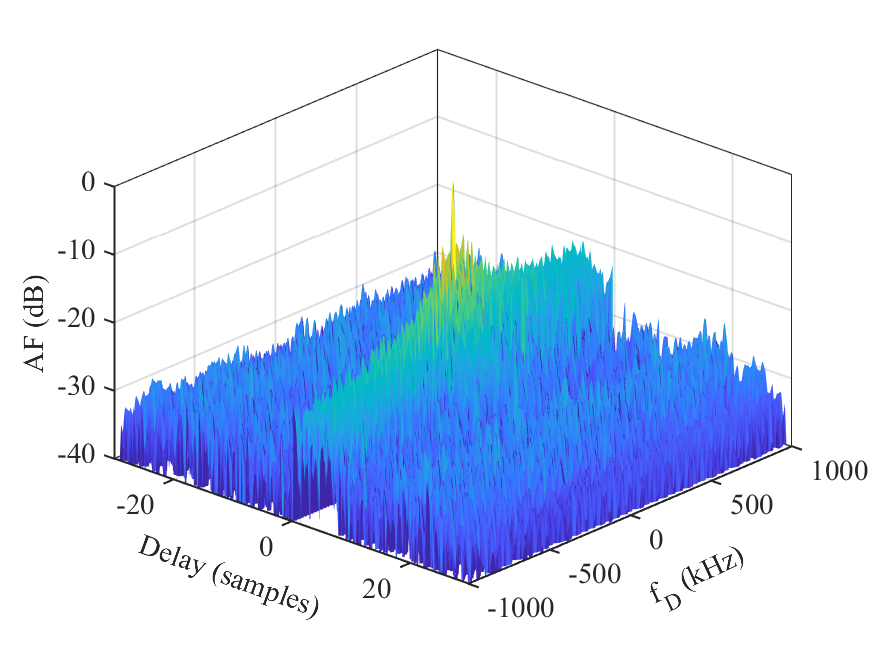}}
  \caption{Three-dimensional delay--Doppler ambiguity surfaces of \ac{PRACH} preambles under large $f_D$ and $\tau$ for (a) \ac{ZC}, (b) \ac{LFM}, and (c) \ac{HFM}.}
  \label{Fig:AF_3D}
\end{figure*}
%%%%%%%%%%%%%%%%%%%%%%%%%%%%%%%%%%%%%%%%%%

%%%%%%%%%%%%%%%%%%%%%%%%%%%%%%%%%%%%%%%%%%
\begin{figure*}
  \subfloat[\label{Fig:correlation}]{
    \includegraphics[width=0.26\linewidth]{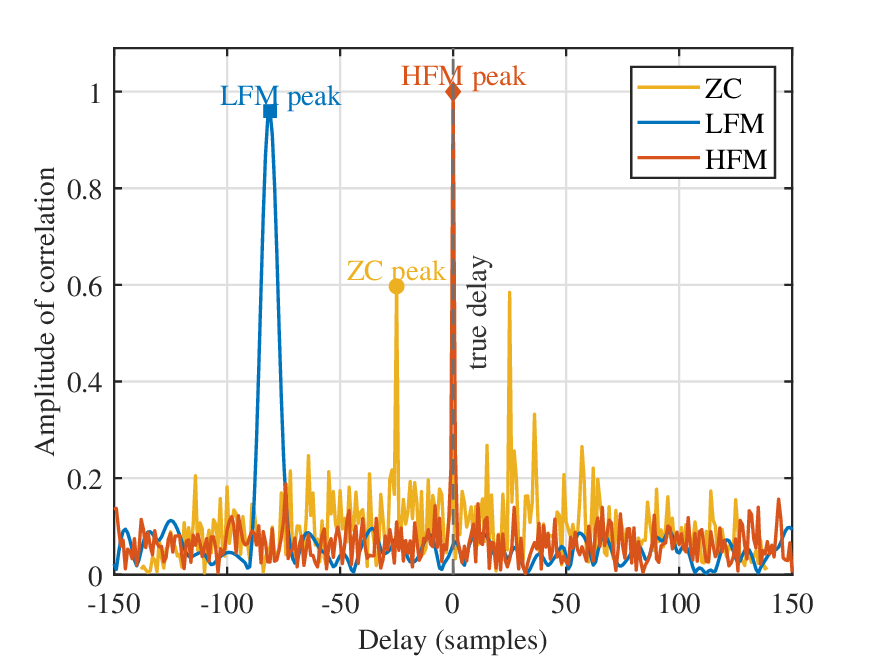}}%
  \hspace{-15pt}
  \subfloat[\label{Fig:ZC_ID}]{
    \includegraphics[width=0.26\linewidth]{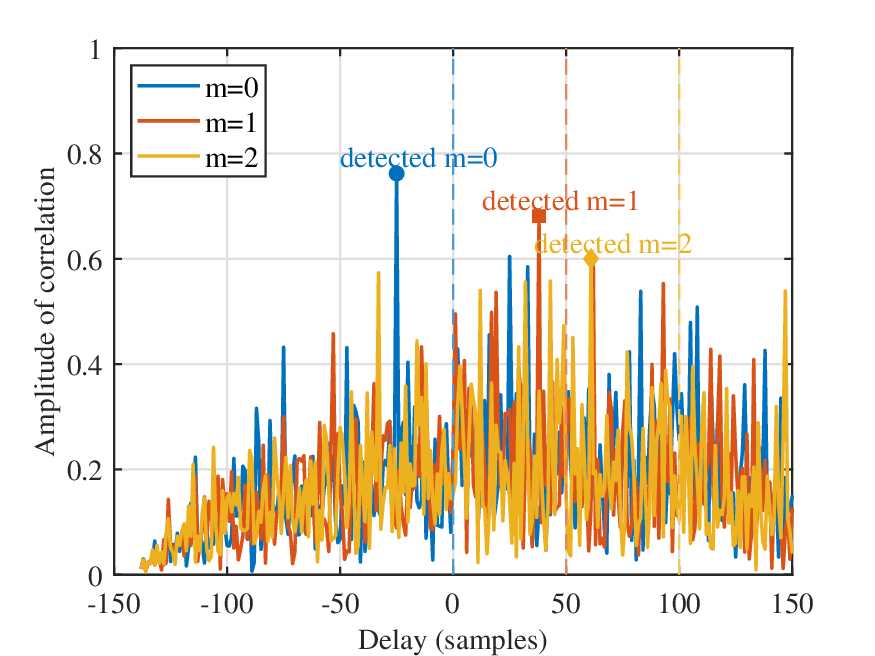}}
  \hspace{-15pt}
  \subfloat[\label{Fig:LFM_ID}]{
    \includegraphics[width=0.26\linewidth]{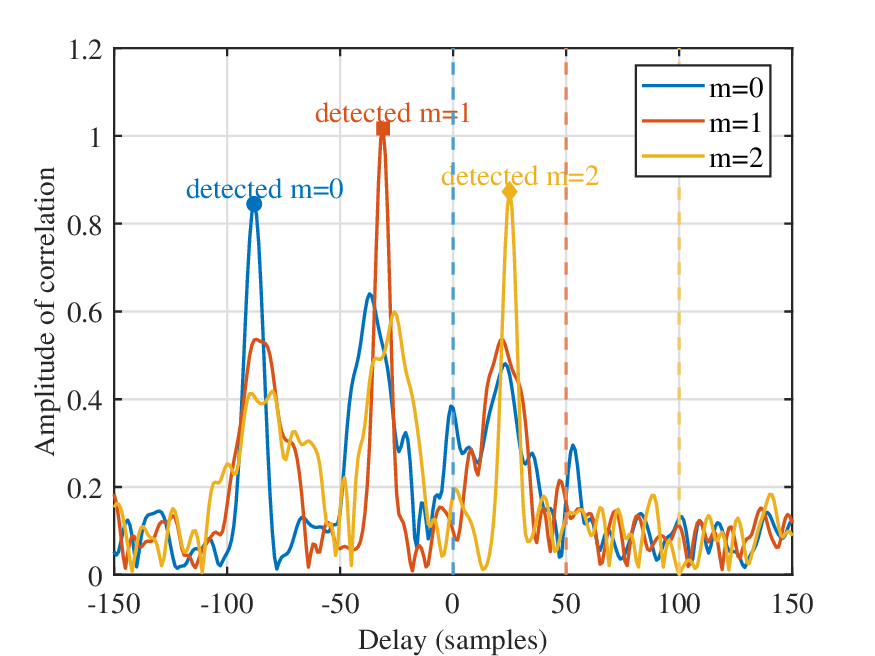}}
  \hspace{-15pt}
  \subfloat[\label{Fig:HFM_ID}]{
    \includegraphics[width=0.26\linewidth]{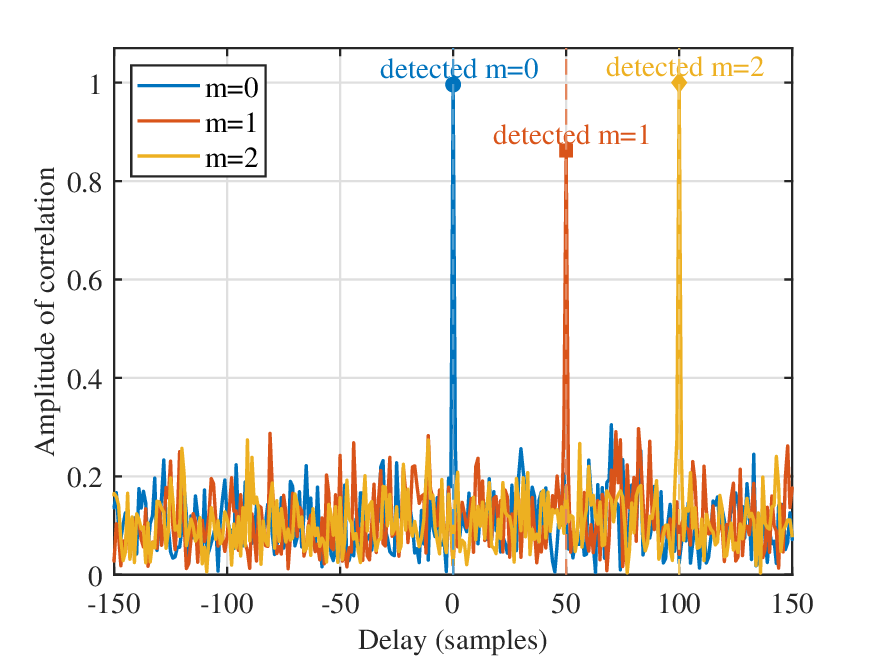}}
  \caption{Correlation behavior and \ac{PID} separability under uncompensated Doppler: (a) single-user timing correlation for \ac{ZC}, \ac{LFM}, and \ac{HFM} at $f_D=500$~kHz with zero timing offset; multi-user \ac{PID} separability for three users at $d_0=0,\, d_1=50,\, d_2=100$ samples, $f_D=300$~kHz, and $\text{\ac{SNR}}=-10$~dB using (b) \ac{ZC}, (c) \ac{LFM}, and (d) \ac{HFM}.}
  \label{Fig:corr_and_ID}
\end{figure*}
%%%%%%%%%%%%%%%%%%%%%%%%%%%%%%%%%%%%%%%%%%

%%%%%%%%%%%%%%%%%%%%%%%%%%%%%%%%%%%%%%%%%%

\section{Simulation Results}
\label{sec:results}
This section assesses the proposed \ac{HFM}-inspired \ac{PRACH} framework through numerical simulations. The detection and identification performance is evaluated under \ac{NTN} scenario.
\subsection{Simulation Setup}
The simulation considers a \ac{LEO}-\ac{NTN} uplink scenario in which the \ac{PRACH} signal experiences a large uncompensated Doppler shift $f_D$ and an unknown timing offset $d$ arising from the one-way propagation delay uncertainty within the \ac{PRACH} observation window. A short-preamble \ac{NR} \ac{PRACH} configuration (Format~A3) is adopted, and the received signal follows the narrowband model in~\eqref{combined_r}. Each \ac{UE} link is modeled as a narrowband Rician fading channel with per-realization gain $h_u = \sqrt{K/(K+1)} + \sqrt{1/(K+1)}\,g_u$, where $g_u \sim \mathcal{CN}(0,1)$ and $K = 10.224$~dB is the Rician factor specified for the \ac{LoS} \ac{NTN} service link in~\cite{3gpp-tr-38.811}. As $K$ is large, the specular component dominates, and detection performance is governed primarily by the Doppler and timing uncertainty. The complete set of system and simulation parameters is provided in Table~\ref{tab:sim_params}.

\subsection{Results Analysis}
This subsection evaluates the timing and multi-user \ac{PID} detection performance of the considered \ac{PRACH} preambles under high-Doppler \ac{NTN} conditions.

\subsubsection{Delay--Doppler Ambiguity Surfaces}
The delay--Doppler surfaces in Fig.~\ref{Fig:AF_3D} confirm the analysis of Section~\ref{Sec:performance_analysis}. The \ac{ZC} self-ambiguity follows a Dirichlet kernel~\eqref{eq:ZC_AF} that, under nonzero Doppler, produces multiple comparable maxima across the delay domain as in Fig.~\ref{Fig:ZC_AF_3D}, yielding non-unique delay estimates. The \ac{LFM} kernel~\eqref{eq:LFM_AF} concentrates along a diagonal ridge set by an affine condition as in Fig.~\ref{Fig:LFM_AF_3D}, shifting the peak systematically with Doppler and introducing a deterministic timing bias. In contrast, the \ac{HFM} delay term~\eqref{eq:HFM_AF} is nonlinear and admits no affine peak condition, precluding ridge or lattice structures; it retains a localized mainlobe at the true delay with no ridge or lattice as in Fig.~\ref{Fig:HFM_AF_3D}, so timing acquisition stays unambiguous across the Doppler search range.
%%%%%%%%%%%%%%%%%%%%%%%%%%%%%%%%%%%%%%%%%%

%%%%%%%%%%%%%%%%%%%%%%%%%%%%%%%%%%%%%%%%%%
\subsubsection{Doppler-Robust Timing Correlation and Multi-PID Detection}
Fig.~\ref{Fig:correlation} evaluates the timing correlation of \ac{ZC}, \ac{LFM}, and \ac{HFM} \ac{PRACH} preambles under a large Doppler shift ($f_D = 500$~kHz) with zero timing offset, isolating the impact of Doppler on the correlation behavior. The proposed \ac{HFM} preamble retains a dominant correlation peak aligned with zero delay, enabling accurate timing estimation without Doppler compensation, consistent with the \ac{AF} analysis in Section~\ref{Sec:performance_analysis}-\ref{sec:ambiguity_analysis}. In contrast, \ac{LFM} produces a well-defined peak that is shifted from the true offset due to its linear delay--Doppler coupling, while \ac{ZC} generates multiple delay locations of comparable amplitude, yielding an ambiguous timing estimate.

A comparative \ac{PID} detection analysis under high-Doppler \ac{NTN} conditions is presented in Fig.~\ref{Fig:corr_and_ID} at $\mathrm{SNR} = -10$~dB, with three simultaneous users at distinct timing offsets ($d_0 = 0,\, d_1 = 50,\, d_2 = 100$ samples) and uncompensated $f_D = 300$~kHz. For \ac{ZC} (Fig.~\ref{Fig:ZC_ID}), Doppler spreads the cross-correlation energy across the delay domain, producing comparable amplitudes at multiple locations that obscure the true delay boundaries and render \ac{PID} detection ambiguous. For \ac{LFM} (Fig.~\ref{Fig:LFM_ID}), the affine coupling generates cross-correlation ridges that leak energy into neighboring \ac{UE} delay regions, increasing inter-user interference despite distinct peaks. In contrast, the proposed \ac{HFM}-inspired \ac{PRACH} (Fig.~\ref{Fig:HFM_ID}) yields well-separated, sharply defined peaks at the true delays: the nonlinear delay--Doppler decoupling preserves per-user timing localization without the Doppler-induced peak migration of \ac{ZC} and \ac{LFM}, enabling robust multi-\ac{PID} identification even under simultaneous access.

%%%%%%%%%%%%%%%%%%%%%%%%%%%%%

%%%%%%%%%%%%%%%%%%%%%%%%%%%%%%%%%%%%%%%%%%
\begin{figure}
\centering
\subfloat[]{%
  \includegraphics[width=0.49\columnwidth]{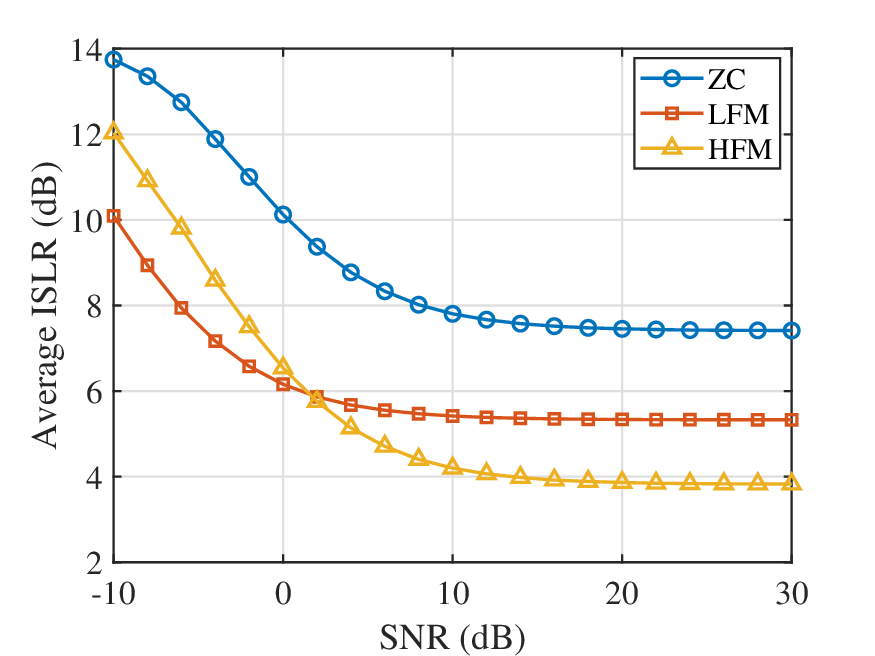}%
  \label{fig:ISLR}}
\hfil
\subfloat[]{%
  \includegraphics[width=0.49\columnwidth]{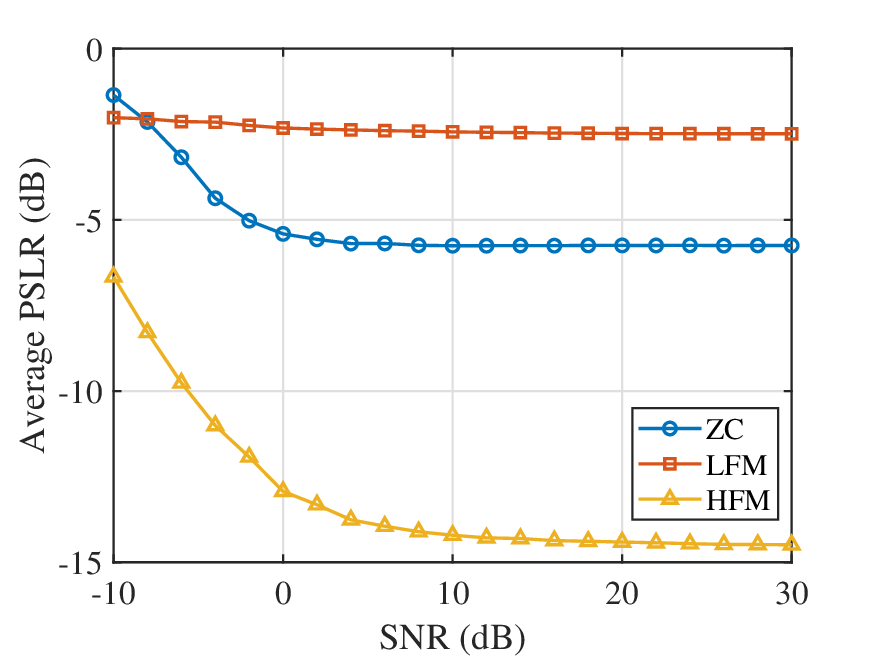}%
  \label{fig:PSLR}}
\caption{Average sidelobe metrics for \ac{ZC}, \ac{LFM}, and \ac{HFM} \ac{PRACH} preambles: (a) \ac{ISLR} and (b) \ac{PSLR}.}
\label{fig:sidelobe}
\end{figure}
%%%%%%%%%%%%%%%%%%%%%%%%%%%%%%%%%%%%%%%%%%

%%%%%%%%%%%%%%%%%%%%%%%%%%%%%%%%%%%%%%%%%%
\subsubsection{ISLR and PSLR Analysis for PRACH Preamble}
The \ac{ISLR} and \ac{PSLR} results in Fig.~\ref{fig:ISLR} and Fig.~\ref{fig:PSLR}, respectively,  compare sidelobe performance across \ac{ZC}, \ac{LFM}, and \ac{HFM} preambles under \ac{NTN} conditions with Doppler and timing uncertainty. Among the considered waveforms, \ac{HFM} achieves the most favorable sidelobe behavior, followed by \ac{LFM}, while \ac{ZC} exhibits the weakest performance, consistent with the \ac{AF} behavior in Fig.~\ref{Fig:AF_3D}. However, in the low-\ac{SNR} region from $-10$ to $0$~dB, \ac{LFM} shows slightly lower \ac{ISLR} (Fig.~\ref{fig:ISLR}) due to energy concentration along its ridge, whereas \ac{HFM}'s distributed peak energy, combined with noise, yields marginally higher sidelobe energy in this regime.

As \ac{SNR} increases, \ac{HFM} achieves lower \ac{ISLR} and significantly improved \ac{PSLR} compared to \ac{LFM} and \ac{ZC}, owing to its well-localized peak and the absence of ridge and lattice structures. The more negative \ac{PSLR} of \ac{HFM} (Fig.~\ref{fig:PSLR}) indicates enhanced main-peak-to-sidelobe separation, leading to more reliable \ac{PID} detection. Both metrics matter for multi-user access; high \ac{ISLR} raises the interference floor and poor \ac{PSLR} weakens peak separation, so \ac{HFM}'s single confined peak yields better separability than \ac{ZC}'s comparable peaks or \ac{LFM}'s ridge leakage.

\subsubsection{PID Detection Performance under Single-User and Multi-User Access}
\begin{figure}
\centering 
\resizebox{0.7\columnwidth}{!}{
\includegraphics{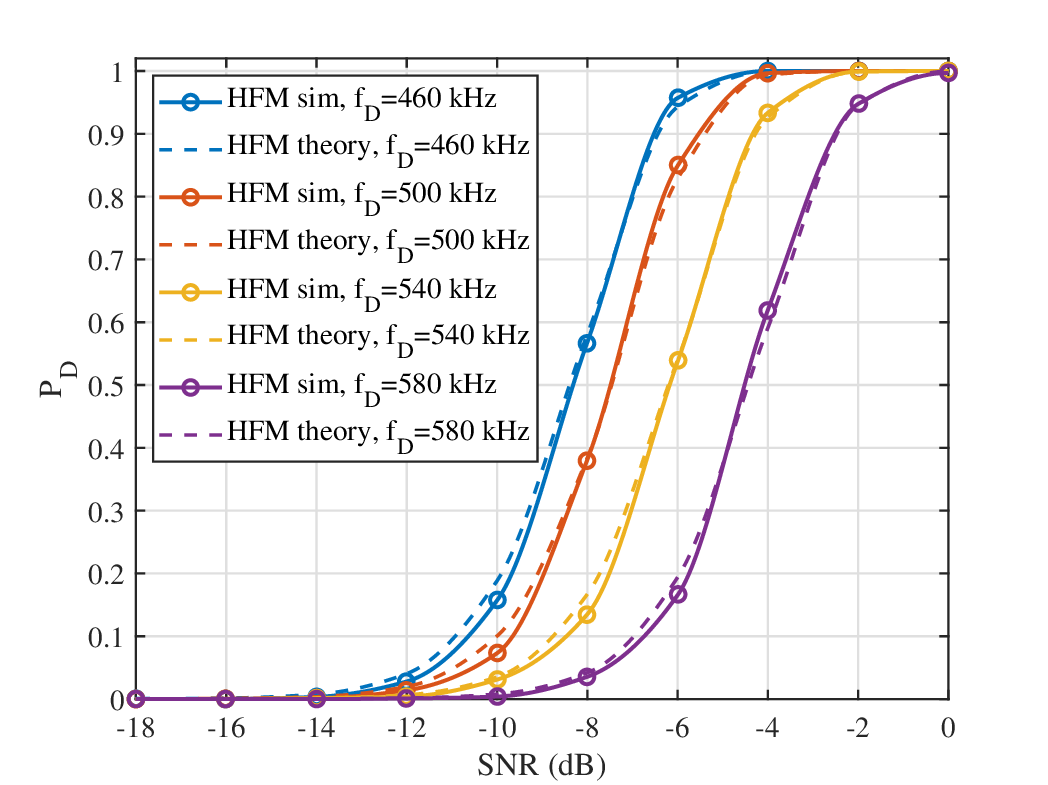}}
\caption{Theoretical versus simulated detection probability $P_D$ of the \ac{HFM} preamble in the single-user case for varying $f_D$.}
\label{fig:hfm_theory}
\end{figure}

\begin{figure*}
  \subfloat[\label{Fig:single_smaller}]{
    \includegraphics[width=0.26\linewidth]{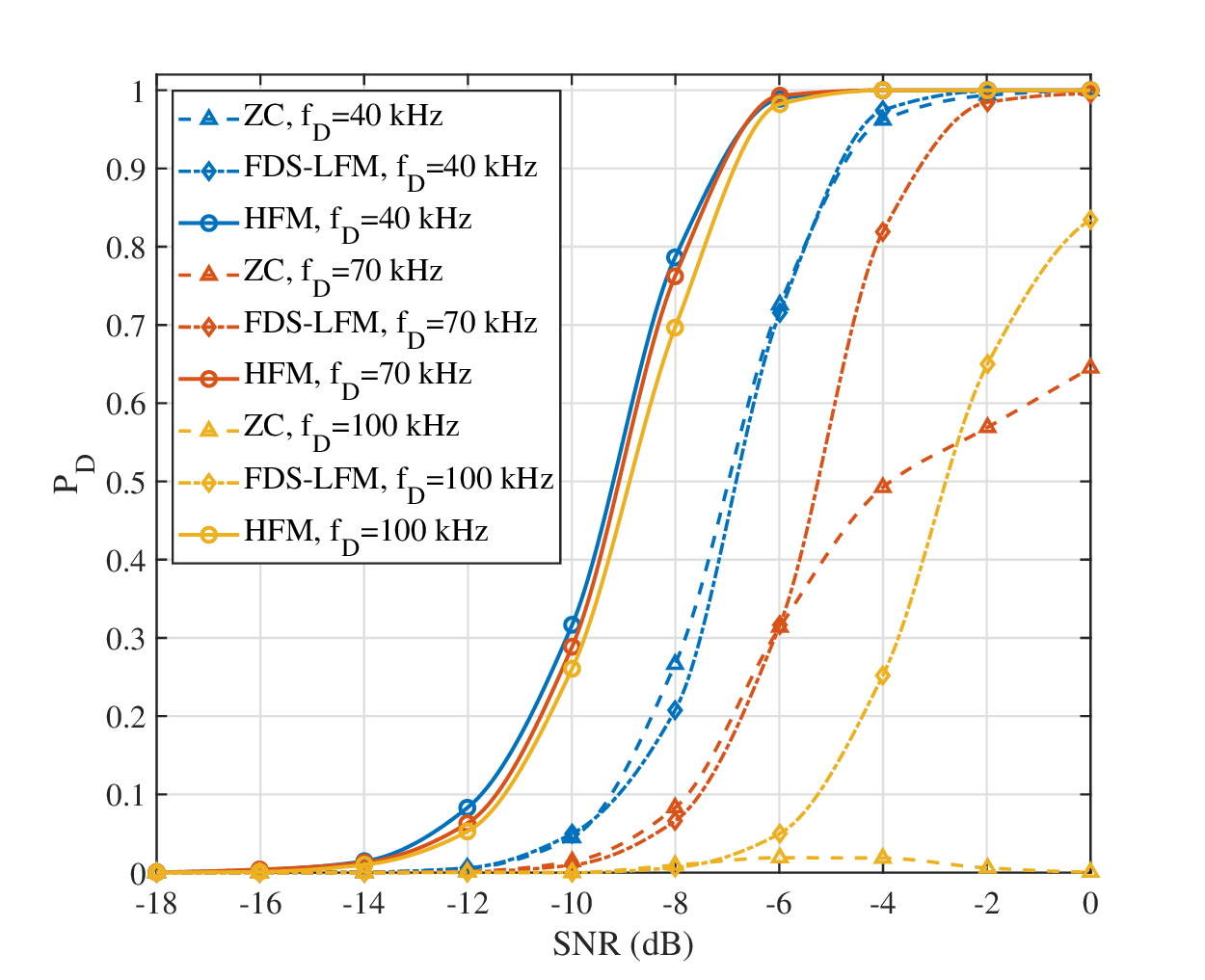}}%
 \hspace{-15pt}
  \subfloat[\label{Fig:single_higher}]{
    \includegraphics[width=0.26\linewidth]{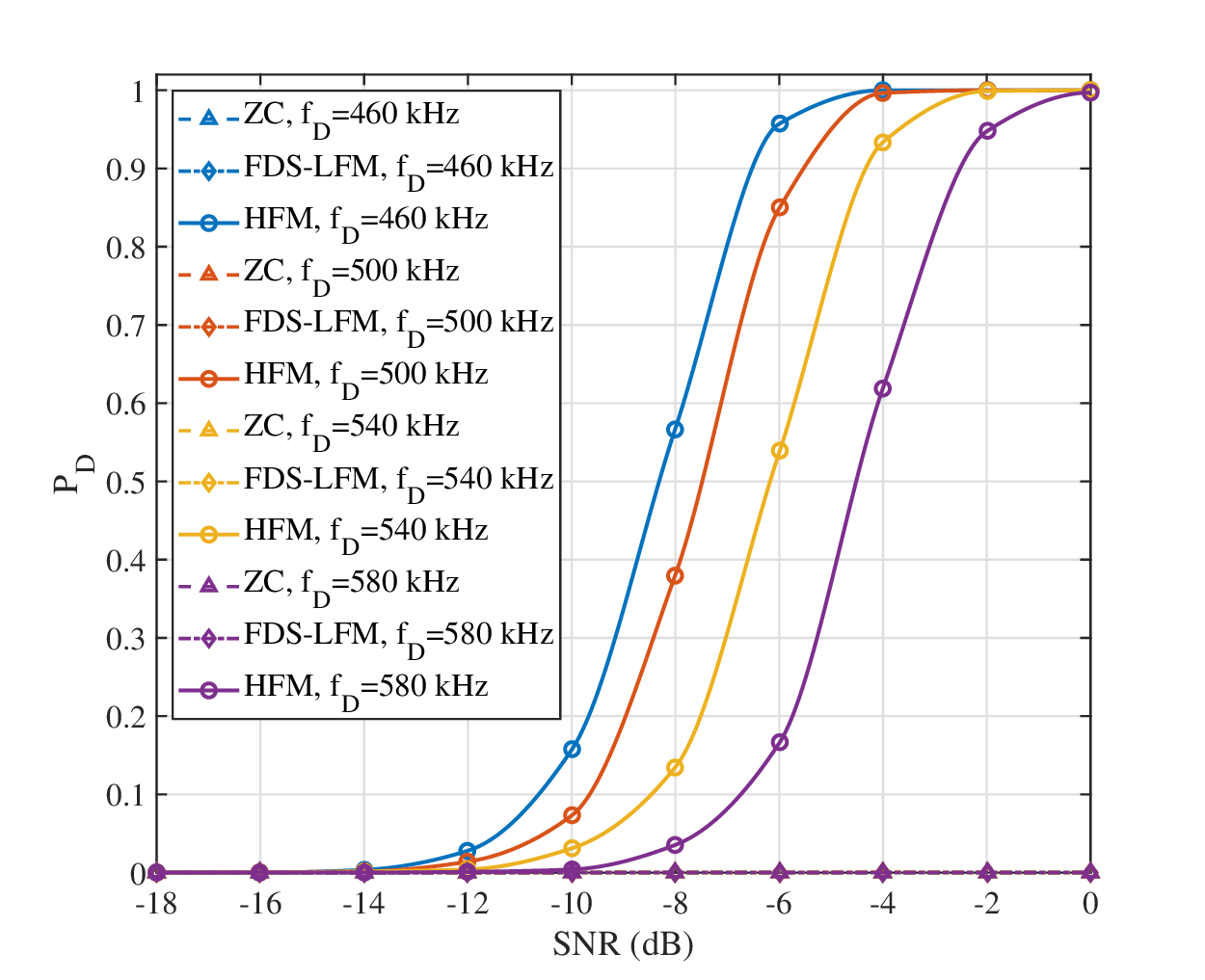}}
     \hspace{-15pt}
\subfloat[\label{Fig:multi_smaller}]{
    \includegraphics[width=0.26\linewidth]{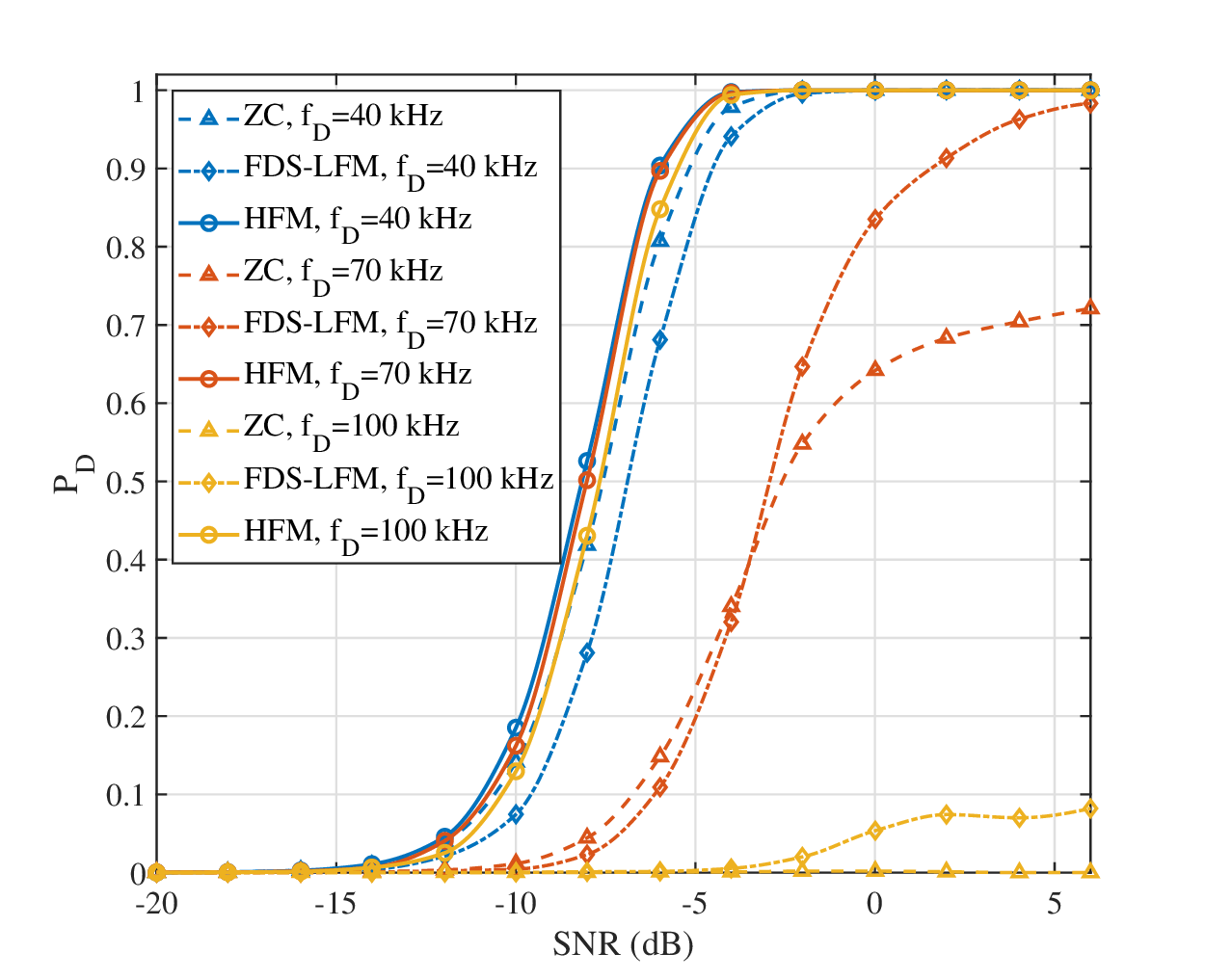}}
   \hspace{-15pt}
  \subfloat[\label{Fig:multi_higher}]{
    \includegraphics[width=0.26\linewidth]{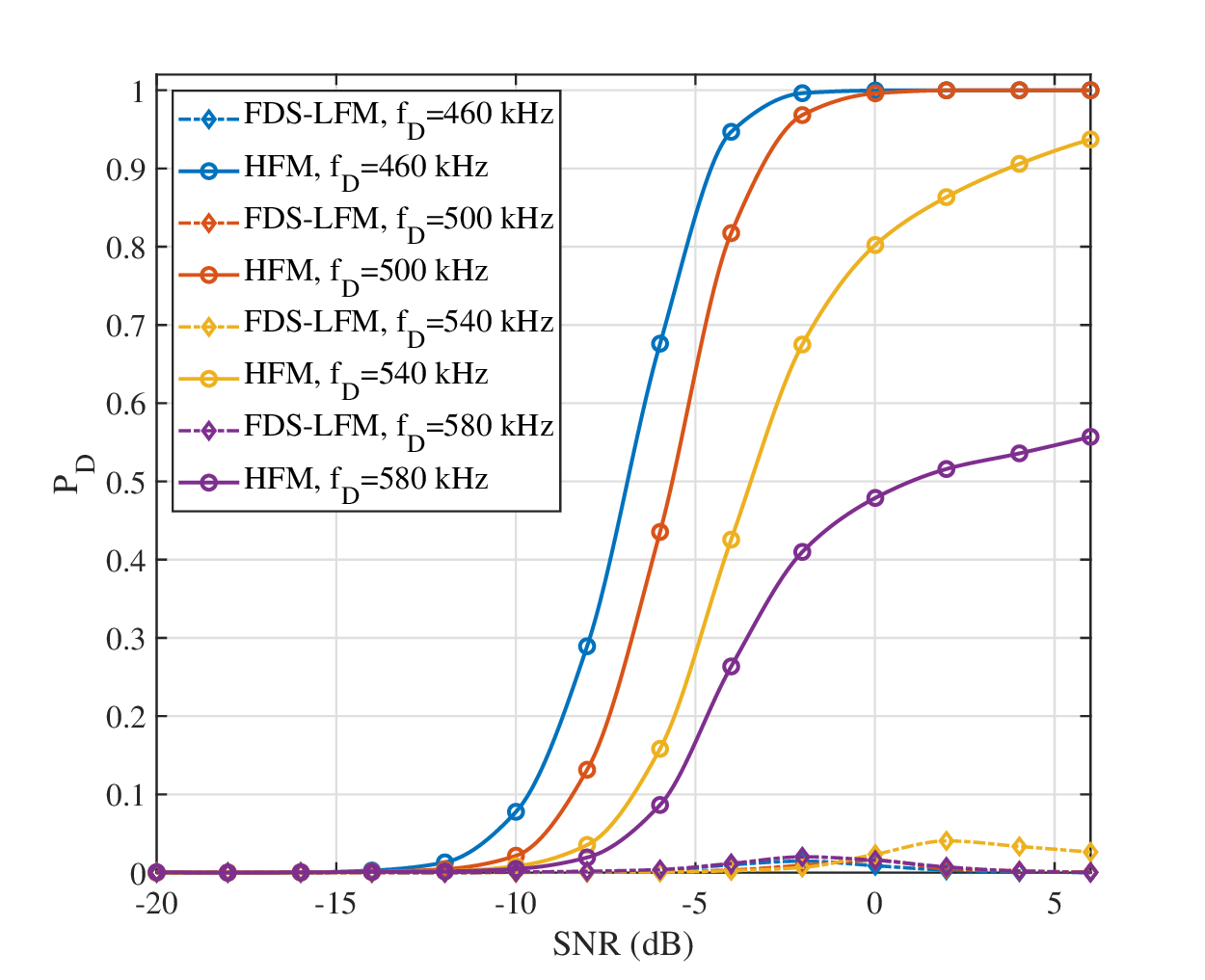}} 
  \caption{Probability of detection for baseline and proposed techniques for (a) lower values of $f_D$ and single-user case (b) higher values of $f_D$ single-user case (c) for lower values of $f_D$ and multi-user case (d) for higher values of $f_D$ and multi-user case.}
  \label{Fig:P_D}
\end{figure*}

Figs.~\ref{fig:hfm_theory} and~\ref{Fig:P_D} show $P_D$ versus \ac{SNR} for varying $f_D$ at a constant $P_{\mathrm{FA}} = 10^{-3}$. The analysis is first validated against simulation, then the three preambles are compared. In Fig.~\ref{fig:hfm_theory}, the analytical \ac{HFM} curves from the first-order Marcum $Q$-function in~\eqref{eq:pd_avg} match the single-user simulation closely across all \ac{SNR} and $f_D$, capturing the transition region and its shift toward higher \ac{SNR} as $f_D$ increases, confirming the accuracy of the analysis.

The \ac{HFM} preamble is then compared against \ac{ZC} and \ac{FDS-LFM} in Fig.~\ref{Fig:P_D}. \ac{FDS-LFM} is used instead of plain \ac{LFM} as it recovers the energy lost by short \ac{LFM} pulses through coherent subsequence combining~\cite{10892192}, though its gain still falls with $f_D$ due to phase misalignment across segments. At low $f_D$ (Fig.~\ref{Fig:single_smaller}), all waveforms detect reliably, with \ac{HFM} reaching a given $P_D$ roughly $2$~dB earlier in \ac{SNR}. At higher $f_D$ (Fig.~\ref{Fig:single_higher}) the baselines degrade sharply, beyond $70$~kHz the \ac{ZC} cyclic correlation breaks down and $P_D \to 0$, and \ac{FDS-LFM} similarly fails as subsequence phase alignment is lost, whereas \ac{HFM} sustains detection since its hyperbolic time--frequency structure decouples delay and Doppler and preserves a dominant correlation peak.

%%%%%%%%%%%%%%%%%%%%%%%%%%%%%%%%%%%%%%%%%%
%%%%%%%%%%%%%%%%%%%%%%%%%%%%%%%%%%%%%%%%%%

%%%%%%%%%%%%%%%%%%%%%%%%%%%%%%%%%%%%%%%%%%
%%%%%%%%%%%%%%%%%%%%%%%%%%%%%%%%%%%%%%%%%%

In the multi-user case (Figs.~\ref{Fig:multi_smaller} and~\ref{Fig:multi_higher}), cross-user interference further exposes the \ac{ZC} and \ac{FDS-LFM} limitations. At $f_D = 40$~kHz all schemes reach $P_D \approx 1$ near $\mathrm{SNR} \approx -6$~dB, but the gap widens with Doppler: \ac{FDS-LFM} needs $2$--$3$~dB more than \ac{HFM} at $70$~kHz and fails to reach high $P_D$ at $100$~kHz, where \ac{ZC} remains near zero. For $f_D \geq 460$~kHz, \ac{FDS-LFM} yields near-zero $P_D$, whereas \ac{HFM} still achieves $P_D \approx 0.9$ at $\mathrm{SNR} \approx -2$~dB, owing to its localized peak structure that enhances separability and suppresses interference.

The narrowband Ricean model adopted in Table~\ref{tab:sim_params} is justified by the operating conditions of the \ac{LEO} Ka-band \ac{NTN} service link. Tropospheric attenuation and ionospheric scintillation vary on timescales substantially longer than the $\sim 50~\mu$s preamble and thus manifest as a slowly varying complex gain absorbed by $h_u$, leaving the phase increments underlying Theorem~\ref{thm:hfm} unaffected. Regarding multipath, the NTN-TDL-C profile in~\cite{3gpp-tr-38.811} reports a \ac{DS} of $[0.21, 6.17]$~ns for Ka-band Rural \ac{LoS}, which lies well below $T_s$; the delayed tap therefore falls within a single sample bin, and since the peak localization of Theorem~\ref{thm:hfm} holds per path, the Doppler invariance of the \ac{HFM} preamble is preserved under standards-compliant operation.

\subsubsection{Timing RMSE Analysis under Single-User and Multi-User Access}
The timing \ac{RMSE} (samples), defined as
%%%%%%%%%%%%%%%%%%%%%%%%%%%%%%%%%%%%%%%%%%
\begin{equation}
\mathrm{RMSE}_d = \sqrt{\mathbb{E}\left[(\hat{d} - d_0)^2\right]},
\end{equation}
%%%%%%%%%%%%%%%%%%%%%%%%%%%%%%%%%%%%%%%%%%
quantifies the accuracy of \ac{PRACH} timing estimation in terms of the sample offset, where $d_0$ and $\hat{d}$ denote the true and estimated
timing offsets, respectively.

%%%%%%%%%%%%%%%%%%%%%%%%%%%%%%%%%%%%%%%%%%
\begin{figure}
\centering 
\resizebox{0.7\columnwidth}{!}{
\includegraphics{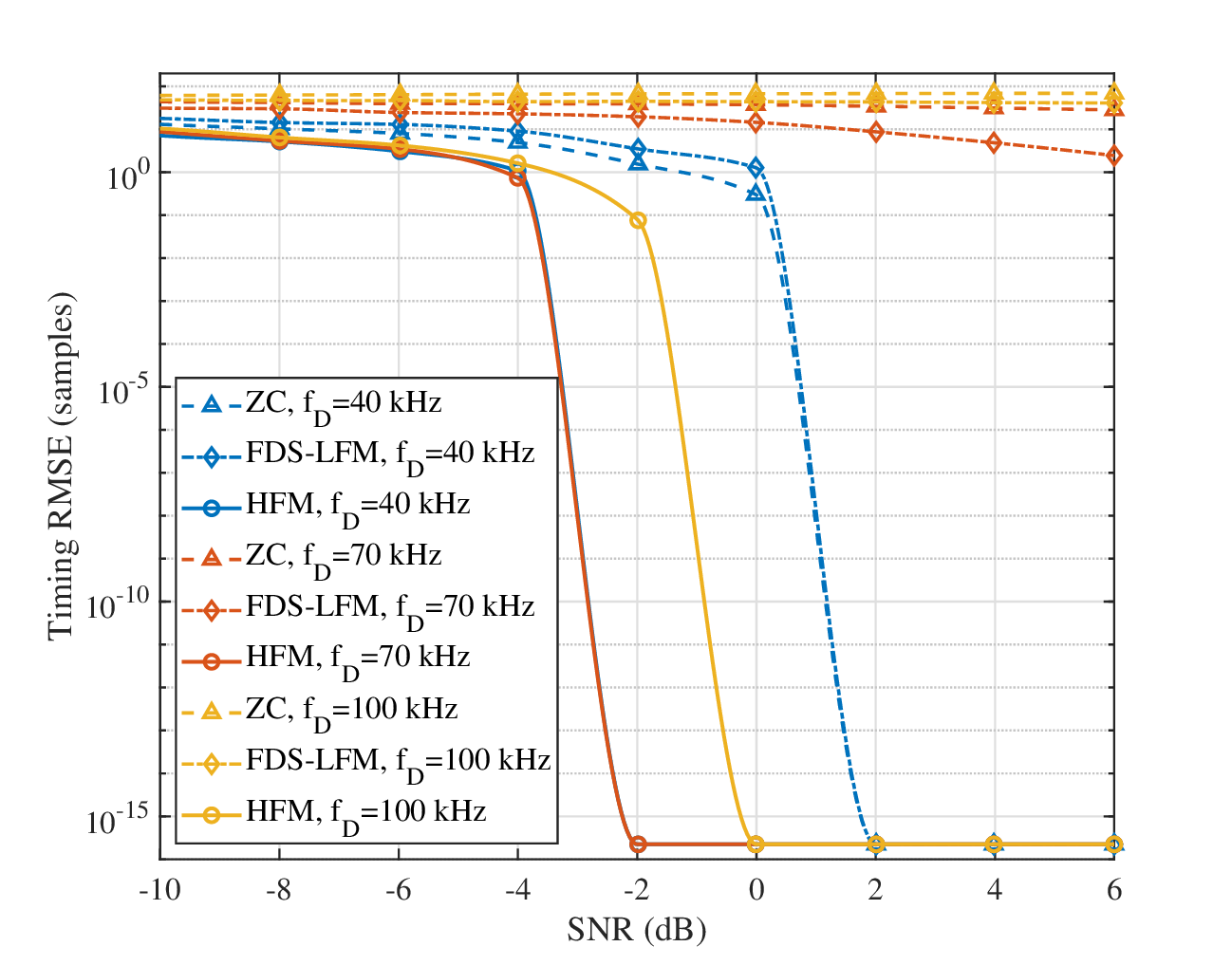}}
\caption{Timing \ac{RMSE} for multi-user case.}
\label{Fig:multi_rmse}
\end{figure}
%%%%%%%%%%%%%%%%%%%%%%%%%%%%%%%%%%%%%%%%%%

In the multi-user case (Fig.~\ref{Fig:multi_rmse}), cross-user interference distorts and overlaps the \ac{ZC} and \ac{FDS-LFM} correlation peaks, raising their \ac{RMSE}. \ac{HFM}, by contrast, preserves a single dominant peak aligned with the true timing instant, as its nonlinear structure mitigates delay--Doppler coupling and cross-user correlation. This yields reduced \ac{RMSE} across $f_D$ and improved timing reliability, consistent with the higher $P_D$ in multi-user scenarios.

%%%%%%%%%%%%%%%%%%%%%%%%%%%%%%%%%%%%%%%%%%
\begin{figure}
\centering 
\resizebox{0.7\columnwidth}{!}{
\includegraphics{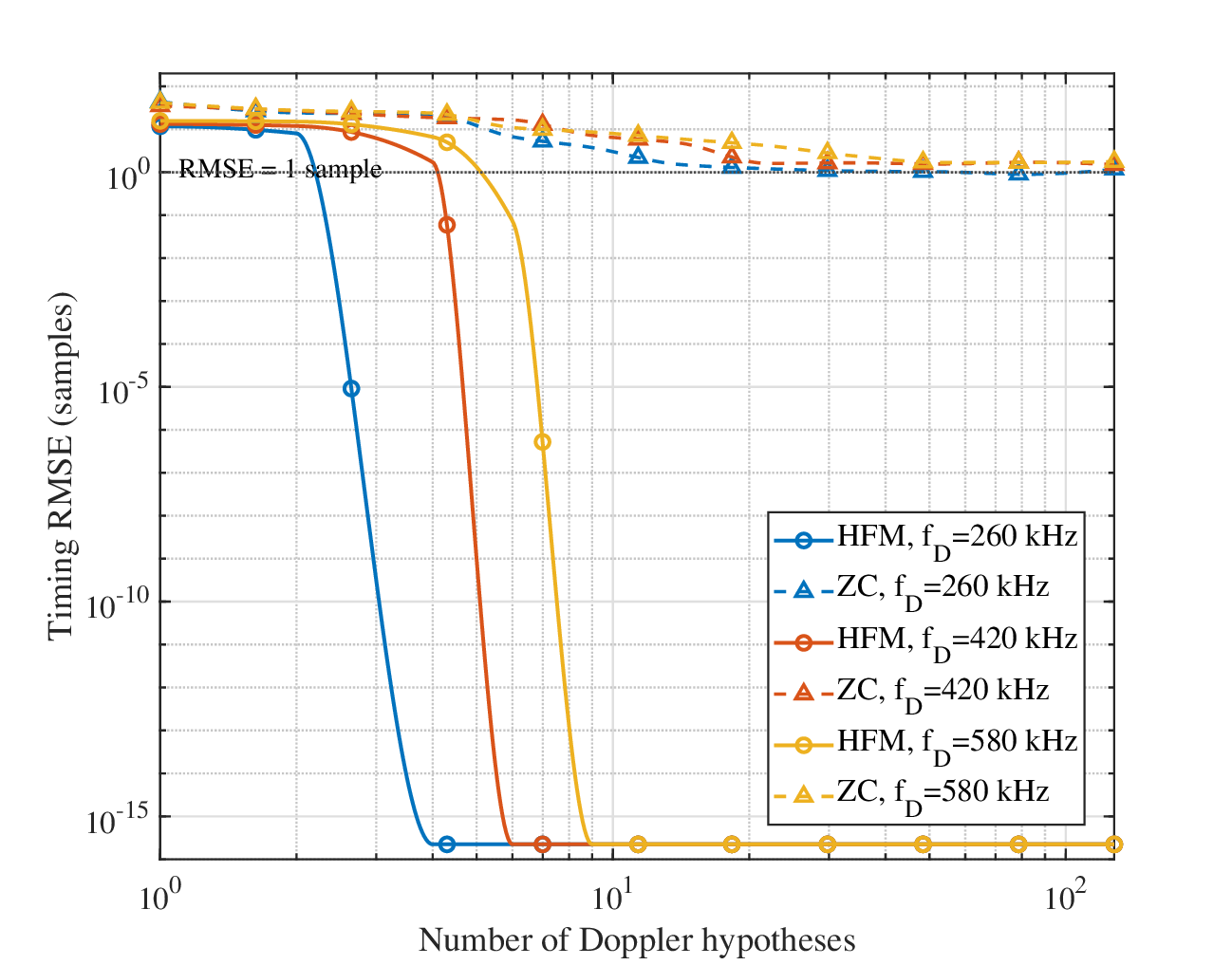}}
\caption{Timing \ac{RMSE} versus number of Doppler hypotheses $|\mathcal{V}|$ for \ac{HFM} and \ac{ZC} preambles at $\ac{SNR}=-6$~dB,
$P_{\mathrm{FA}}=10^{-3}$.}
\label{Fig:doppler_bin_rmse}
\end{figure}
%%%%%%%%%%%%%%%%%%%%%%%%%%%%%%%%%%%%%%%%%%

Fig.~\ref{Fig:doppler_bin_rmse} shows the timing \ac{RMSE} versus the number of Doppler hypotheses $|\mathcal{V}|$ at $\ac{SNR}=-6$~dB and $P_{\mathrm{FA}}=10^{-3}$, with the threshold recalibrated per $|\mathcal{V}|$. Under the one-sample criterion, \ac{HFM} reaches sub-sample accuracy with only a few hypotheses, whereas \ac{ZC} stays above one sample across the swept range, empirically validating $|\mathcal{V}_{\text{ZC}}| \gg |\mathcal{V}_{\text{HFM}}|$ from Section~\ref{Sec:performance_analysis}-\ref{subsec:compexity}. This disparity arises from two mechanisms. First, the \ac{HFM} delay response does not migrate with Doppler; the peak stays at the true lag and is only attenuated by the residual offset, so a small Doppler bank suffices, whereas \ac{ZC} exhibits Doppler-induced peak migration that necessitates a considerably denser grid. Second, at high $f_D$ the \ac{ZC} \ac{AF} produces several peaks of comparable magnitude, so that the dominant peak does not necessarily coincide with the true timing instant, sustaining the \ac{RMSE} above the threshold as $|\mathcal{V}|$ increases.

%%%%%%%%%%%%%%%%%%%%%%%%%%%%%%%%%%%%%%%%%%
\section{Conclusion}
\label{sec:conclusion}
This paper addressed reliable random access in \ac{LEO}-based \acp{NTN}, where large uncompensated Doppler and significant delay uncertainty undermine the timing estimation and preamble identification of conventional \ac{NR} \ac{PRACH} designs. Through a unified delay–Doppler ambiguity function framework, we showed that \ac{ZC} and \ac{LFM} preambles are fundamentally limited by lattice and ridge structures that couple delays with Doppler. To overcome these limitations, an \ac{HFM}-inspired \ac{PRACH} preamble with a scaling-factor-parameterized codebook was proposed, fully compatible with the existing \ac{NR} \ac{PRACH} structure. Analytical and simulation results confirmed that the proposed design suppresses lattice and ridge structures, thereby enabling timing estimation free of deterministic Doppler-dependent bias and reliable preamble identification at large uncompensated $f_D$ while consistently outperforming the \ac{ZC} and \ac{FDS-LFM} baselines under severe Doppler. 
%%%%%%%%%%%%%%%%%%%%%%%%%%%%%


\end{document}